\pdfoutput=1 
\documentclass[namedreferences]{SolarPhysics}
\usepackage[optionalrh,solaenum]{spr-sola-addons} 
\usepackage{graphicx}        
\usepackage{color}           
\usepackage{url}             

\newcommand{\be}{\begin{equation}}
\newcommand{\ee}{\end{equation}}

\newcommand{\anngeo}{   {\it Annales Geophysicae}}

\newcommand{\aap}{    {\it Astron. Astrophys.}}
\newcommand{\aaps}{   {\it Astron. Astrophys. Suppl.}}

\newcommand{\apj}{    {\it Astrophys. J.}}
\newcommand{\apjl}{{\it Astrophys. J. Lett.}}
\newcommand{\memit}{{\it Mem. della Soc. Astr. Ital.}}

\newcommand{\grl}{    {\it Geophys. Res. Lett.}}

\newcommand{\jgr}{    {\it J. Geophys. Res.}}
\newcommand{\jgrsp}{    {\it J. Geophys. Res. Sp.}}

\newcommand{\solphys}{{\it Solar Phys.}}

\newcommand{\basi}{    {\it  Bul. of the Astron. Soc. of India.}}

\begin{document}

\begin{article}

\begin{opening}

\title{Predictions of energy and helicity in four major eruptive solar flares}

\author{M.~\surname{Kazachenko}$^{1}$\sep
        R.~\surname{Canfield}$^{1}$\sep
        D.~\surname{Longcope}$^{1}$\sep
        J.~\surname{Qiu}$^{1}$}
\runningauthor{Kazachenko et al.}
\runningtitle{Predictions of energy and helicity in four major eruptive solar flares}

   \institute{$^{1}$ Montana State University
                     email: \url{kazachenko@physics.montana.edu} \\
              }

\begin{abstract}
In order to better understand the solar genesis of interplanetary magnetic clouds (MCs) we model the magnetic and topological properties of four large eruptive solar flares and relate them to observations. We use the three-dimensional Minimum Current Corona model \cite{Longcope1996d} and observations of pre-flare photospheric magnetic field and flare ribbons to derive values of reconnected magnetic flux, flare energy, flux rope helicity and orientation of the flux rope poloidal field. We compare model predictions of those quantities to flare and MC observations and within the estimated uncertainties of the methods used find the following. The predicted model reconnection fluxes are equal to or lower than the reconnection fluxes inferred from the observed ribbon motions. Both observed and model reconnection fluxes match the MC poloidal fluxes. The predicted flux rope helicities match the MC helicities. The predicted free energies lie between the observed energies and the estimated total flare luminosities. The direction of the leading edge of the MC's poloidal field is aligned with the poloidal field of the flux rope in the AR rather than the global dipole field. These findings compel us to believe that magnetic clouds associated with these four solar flares are formed by low-corona magnetic reconnection during the eruption, rather than eruption of pre-existing structures in the corona or formation in the upper corona with participation of the global magnetic field. We also note that since all four flares occurred in active regions without significant pre-flare flux emergence and cancellation, the energy and helicity we find are stored by shearing and rotating motions, which are sufficient to account for the observed radiative flare energy and MC helicity.
\end{abstract}
\end{opening}

\section{Introduction} \label{secintro4}

Coronal mass ejections (CMEs) expel plasma, magnetic flux and helicity from the Sun into the interplanetary medium. At 1 AU in the interplanetary medium CMEs appear
as interplanetary coronal mass ejections (ICMEs). At least one third \cite{Gosling1990} or perhaps a larger fraction \cite{Webb.etal2000_rcc} of the ICMEs observed in situ are magnetic clouds (MCs) \cite{Burlaga1981}, coherent ``flux-rope'' structures characterized by low proton temperature and strong magnetic field whose direction typically rotates smoothly as they
pass the satellite.

MCs originate from eruptions of both quiescent filaments and active regions (ARs). The 3D magnetic models and geomagnetic relationships are better understood for filament eruptions than for
ARs \cite{Marubashi1986,Bothmer1998,Zhao1998,Yurchyshyn2001}. However the most geoeffective MCs originate from ARs \cite{Gopalswamy2010}. In this paper
we focus exclusively on the latter.

Comparison of the properties of MCs with those of their related ARs clarifies our understanding of both domains.
Assuming MCs to be twisted flux ropes in magnetic equilibrium, several authors have succeeded in inferring global properties such as MC axis orientation, net magnetic flux,
and  magnetic helicity (see review by \inlinecite{Demoulin2008}). For a sample of twelve MCs \inlinecite{Leamon2004}  found that the percentage of MC poloidal flux relative to unsigned vertical AR flux
varied widely, from $1\%$  to $~300\%$. For one MC \inlinecite{Luoni2005}  did a similar study and found a factor 10 times lower flux in the MC than the AR, in agreement
with other previous studies \cite{Demoulin2002,Green2002}. More recently, for a sample of nine MCs, \inlinecite{Qiu2007}
found that the MC poloidal flux matches the ``observed'' reconnection flux, i.e. reconnection flux in the two-ribbon
flare associated with it, and the toroidal flux is a fraction
of the reconnection flux. The \inlinecite{Qiu2007} results may be interpreted as evidence
of formation of the helical structure of magnetic flux ropes by reconnection, in the course of which magnetic flux, as well as helicity, is transported into the flux rope.

The other quantity that is very useful for relating MCs to their associated flares is magnetic helicity, which describes how sheared and twisted the magnetic field is compared to its
lowest energy state \cite{Berger1999,Demoulin2007}. Since helicity is approximately conserved in the solar atmosphere and the heliosphere \cite{Berger1984}, it is a very powerful quantity for linking
solar and interplanetary phenomena. For six ARs \inlinecite{Nindos2003} found that photospheric helicity injection in the whole AR is comparable with the MC helicity. However, it is worth remembering that this approach is simplified, since the liftoff of the flux rope does not remove all of the helicity available in the AR \cite{Mackay2006,Gibson2008}. \inlinecite{Mandrini2005} and \inlinecite{Luoni2005} compared, respectively, the helicity released from a very small AR and a very large AR, with the helicity  of their associated MC. They found a very good agreement in the values (small AR with small MC, and large AR with large MC), despite a difference of $3$ orders of magnitude between the smaller and the larger events.

There exist two basic ideas about the solar origin of magnetic clouds: MCs are formed either globally or locally. In the {\it global} picture, the MC topology
is defined by the overall dipolar magnetic field of the Sun \cite{Crooker2000_rcc}.  In this case, the field lines of the helmet streamer belt become the outermost coils
 of the MC through reconnection behind the CME as it lifts off.  Hence the leading field direction of magnetic cloud tends to follow that of  the large-scale solar dipole,
  reversing at solar maximum  \cite{Mulligan1998,Li2010,Bothmer1998}. In the {\it local} picture, on the other hand, the flux rope is formed within the AR and its properties are defined by
   properties of the AR. We can categorize the ``local'' models into two sub-classess. In the first, the magnetic flux rope emerges from beneath the photosphere into the corona
   \cite{Low1994,Chen1989,Fan2004,Leka1996,Abbett2003}. In this scenario the flux ropes formed  may maintain stability for a relatively long time
   prior to the explosive loss of equilibrium \cite{Forbes1995,Lin2004} or a breakout type reconnection that opens up the overlying flux rope of opposite polarities \cite{Antiochos1999}.
Such a flux rope is therefore {\it pre-existing} before its expulsion into interplanetary space. In the second case the flux rope is formed {\it in situ} by magnetic reconnection.
The magnetic reconnection suddenly re-organizes the field configuration in favor of expulsion of the ``in situ'' formed magnetic flux rope out of the solar atmosphere. The results
of \inlinecite{Qiu2007} support this case. \inlinecite{Qiu2007} found that the reconnection flux from observations of flare ribbon evolution
is greater than toroidal flux of the MC but comparable and
proportional to its poloidal flux, regardless of the presence of filament eruption. Their conclusion agrees with the inference from the study by \inlinecite{Leamon2004}, although through
a very different approach.

Our working hypothesis is that MCs associated with the ARs originate from the ejection of locally in-situ formed flux ropes. In this case shearing and rotation of the photosphere magnetic flux concentration before the flare lead to the buildup of magnetic stress which is removed during the flare by reconnection. As a result a magnetic flux rope is formed and erupts, producing a MC.

To test our hypothesis we  apply a quantitative non-potential self-consistent model, the  \emph{Minimum Current Corona} (MCC) model \cite{Longcope1996d,Longcope2001b}, to predict the properties of the in-situ formed flux rope in four two-ribbon flares. Using the MCC model with SOHO/MDI magnetogram sequences we construct a three-dimensional model of the
 pre-flare magnetic field topology and make quantitative predictions of the amount of magnetic flux that reconnects in the flare, the magnetic self-helicity of the flux rope
 created, and the minimum energy release the topological change would yield. We then compare the predicted flare helicity and energy to MC helicity and flare energy, inferred from fitting the magnetic cloud (Wind/ACE) and GOES observations correspondingly. We compare the predicted reconnected flux to the amount of photospheric flux swept up
  by the flare ribbons using TRACE 1600 \AA{} data and the poloidal MC flux inferred from fitting the magnetic cloud observations.   We find that for the four studied flares our results support, from the point of view of flux, energy and helicity, the scenario in which the MC progenitor is a helical flux rope formed in situ by magnetic reconnection in the low corona immediately before its expulsion into interplanetary space. We also find that MC topology is defined by the local AR structure rather than the overall dipolar magnetic field of the Sun in the events studied.

This paper is organized as follows: in Section~\ref{methods4} we describe the methods and uncertainties of our analysis.  In Section~\ref{flares4} we describe
the four flares studied, the flux and helicity of the ARs in which the flares occurred, and the magnetogram sequence during the buildup time.
In Section~\ref{discussion4} we discuss our results, and in Section~\ref{conclusion} summarize our conclusions.

\section{Methods: Calculating Reconnection Flux, Energy and Helicity}\label{methods4}
In this section we describe the methods that we use to (\S\ref{MCC4}) predict the reconnection flux, energy and helicity from SOHO/MDI magnetogram sequences and the Minimum Current
Corona model and (\S\ref{obs4}) determine observed values of these quantities from GOES, TRACE, ACE, and WIND observations.

\subsection{Minimum Current Corona Model}\label{MCC4}
The key improvement of our study relative to \inlinecite{Leamon2004} and \inlinecite{Qiu2007} is the use of the \emph{Minimum Current Corona} model, which allows us to estimate the energy
and helicity associated with the in-situ formed flux rope \cite{Longcope1996d,Longcope2001b}.
The MCC model extends the basic elements of the CSHKP \cite{carmichael1964,Sturrock1968,Hirayama1974,Kopp1976} two-ribbon flare scenario to three dimensions, including the shearing of an AR
along its polarity inversion line (PIL) to build up stress. After this pre-flare stress buildup, the MCC model quantifies the result of eliminating some or all of the stress
and creating a twisted flux rope overlying the AR, through magnetic reconnection.

\begin{table*} \caption{\small Flare and AR Properties (see \S\ref{flares4}).  Number is the NOAA number of the AR associated with the flare, $\Phi_{AR}$, in units of  $10^{22}$ Mx, is the AR's unsigned magnetic flux and $H_{AR}$, in units of $10^{42}$ Mx$^{2}$, is helicity injected into the AR during the magnetogram sequence starting at $t_{0}$ and ending at $t_{flare}$. } \begin{center}
\begin{tabular}{c|ccc|ccc|c|}
 $i$ & \multicolumn{3}{|c|}{Flare} & \multicolumn{3}{|c|} {Active Region} & {M-gram sequence} \\
 &  Date & Time & Class & Number & $\Phi_{AR}$ & $H_{AR}$ & $t_0$~~~~~~~~$t_{flare}$ \\
\hline
1 & 05/13 2005 & 16:57 & M8 & 10759& $ 2.0$ & $-12\pm1.2$ & \multicolumn{1}{|l|}{05/11 23:59} \\
& & & & & & & \multicolumn{1}{|r|}{05/13 16:03} \\
2 & 11/07 2004 & 16:06 & X2 & 10696& $ 2.1$ & $-15\pm1.5$ & \multicolumn{1}{|l|}{11/06 00:03} \\
& & & & & & & \multicolumn{1}{|r|}{11/07 16:03}  \\
3 & 07/14 2000 & 10:03 & X6 & 09077& $ 3.4$ & $-27\pm2.7$ & \multicolumn{1}{|l|}{07/12 14:27} \\
& & & & & & & \multicolumn{1}{|r|}{07/14 09:36}  \\
4 & 10/28 2003 & 11:10 & X17 & 10486& $ 7.5$ & $-140\pm14.$& \multicolumn{1}{|l|}{10/26 12:00} \\
& & & & & & & \multicolumn{1}{|r|}{10/28 10:00}  \\
\end{tabular}
\normalsize \label{tab:t01_4} \end{center} \end{table*}

To describe the evolution of the pre-flare photospheric motions that lead to stress build-up we use a sequence of SOI/MDI full-disk magnetograms \cite{Scherrer1995}. As the starting point we take $t_0$, right after the end of a large flare, which we call the zero-flare.  We assume that at $t_0$ the AR's  magnetic field becomes fully relaxed. As the ending time we take $t_{flare}$, right before the time when the flare of study occurred but avoiding artifacts associated with the onset of the flare brightening
\cite{Qiu2003}.  To achieve the maximum energy release, the field reconnecting
during the flare would need to relax to its potential state,
hence we assume the field to be potential at $t_{flare}$. As a result, we form a sequence of magnetograms, which covers $\Delta t$ hours of stress build-up prior to the flare (see Table~\ref{tab:t01_4}).

For quantitative analysis of the pre-flare magnetic field we divide each magnetogram into a set of unipolar {\it partitions} and then into unipolar magnetic charges
(e.g. see partitioned magnetogram in Appendix, Figure~\ref{fig:f04_4}). Firstly, for all successive pairs of magnetograms we derive a local correlation tracking (LCT)
velocity field \cite{November1988,Chae2001} and then group pixels into individual partitions that have persistent labels.  In the second step we represent each
magnetic partition with a magnetic {\it point charge} (or magnetic point source) which has the flux of the partition and is located at its center of flux.
Finally, using the LCT velocity field we calculate the helicity injected by the motions of the magnetic point  charges of the whole AR, ($H_{AR}$, see Table~\ref{tab:t01_4} and \inlinecite{Longcope2007}).
 We make sure that the amount of helicity injected by the motions of the continuous photospheric partitions matches the helicity injected by the motions of the magnetic
point charges. Their equality gives us confidence that the centroid motions of the point charges accurately capture helicity injection.
Computing the vector potential via the Fourier approach of \inlinecite{Chae2001}, as we choose to do, results in the higher values (10\%) in the helicity flux compared to that
from the approach by \inlinecite{Pariat2005} \cite{Chae2007}.  In addition, the LCT method that we use yields systematically lower values than
the DAVE velocity inversion algorithm with a difference in helicity flux of less than $\simeq10\%$ \cite{Welsch2007}. Those two effects result in an uncertainty of $10\%$ in the $H_{AR}$ value which we take into account.

\begin{table*} \caption{\small Flare physical properties: MCC model predictions vs. observations:
predicted ($\Phi_{r,MCC}$) and inferred from observations ($\Phi_{r,ribbon}$) reconnection fluxes and MC poloidal fluxes ($\Phi_{p,MC}$), predicted
($\mathcal{E}_{MCC}$) and observed ($\mathcal{E}_{GOES}$) energy values, predicted ($H_{MCC}$) and observed ($H_{MC}$) helicity values (see \S\ref{flares4}).
} \begin{center}
\begin{tabular}{c|ccc|c|c|}
$i $  & $\Phi_{r,MCC}$ & 	$\Phi_{r,ribbon}$ & $\Phi_{p,MC}$ & $\mathcal{E}_{MCC}$~~ $\mathcal{E}_{GOES}$  & $H_{MCC}$~~$H_{MC}$\\
 &  $10^{21}$ Mx & $10^{21}$ Mx  &  $10^{21}$  Mx & $10^{31}$ ergs~~$10^{31}$ ergs &  $10^{42}$ Mx$^{2}$~~$10^{42}$ Mx$^{2}$   \\
\hline
1  & $2.8\pm0.4$  & $4.1\pm0.4$ & $6.3\pm4.2$ & \multicolumn{1}{|l|}{$1.0\pm0.3$} & \multicolumn{1}{|l|}{$-7.0\pm1.2$} \\
   &&&& \multicolumn{1}{|r|}{$3.1\pm0.6$} & \multicolumn{1}{|r|}{$-7.5\pm5.0$} \\
2  & $5.4\pm0.8$  & $4.8\pm0.5$ & $5.25\pm3.5$ & \multicolumn{1}{|l|}{$6.4\pm1.8$} & \multicolumn{1}{|l|}{$-5.0\pm0.6$}\\
   &&&&\multicolumn{1}{|r|}{$2.0\pm0.1$} & \multicolumn{1}{|r|}{$-8.3\pm5.5$} \\
3  & $6.0\pm0.9$  & $12.8\pm3$  & $9.9\pm6.6$ & \multicolumn{1}{|l|}{$9.1\pm2.6$} & \multicolumn{1}{|l|}{$-20.1\pm3.6$}\\
   &&&&\multicolumn{1}{|r|}{$10.1\pm2.1$} & \multicolumn{1}{|r|}{$-22.5\pm15.0$} \\
4  & $15.0\pm2.6$ & $23\pm7$    & $18.0\pm12.0$ & \multicolumn{1}{|l|}{$18.0\pm5.2$} & \multicolumn{1}{|l|}{$-48.0\pm8.6$}\\
   &&&&\multicolumn{1}{|r|}{$13.6\pm0.6$} & \multicolumn{1}{|r|}{$-45.0\pm30.0$} \\
\end{tabular}
\normalsize \label{tab:t02_4} \end{center} \end{table*}

The MCC model characterizes the changes in the pre-flare magnetic field purely in terms of the changes in the {\it magnetic domains}, volumes of field lines connecting
pairs of opposite point charges. Replacing each partition with a single magnetic point
charge as we chose to do results in values of domain fluxes
that are only slightly different from the actual domain fluxes
\cite{Longcope2009}. As magnetic charges move, the magnetic field, first relaxed by the zero-flare,  becomes increasingly stressed and hence non-potential.
Under the assumption that no reconnection, flux emergence or cancellation occur between the zero-flare and the flare of interest, the domain fluxes could not have changed. (Note, that for our analysis we selected only
flares associated with ARs with no significant flux emergence
or cancellation during the period between $t_0$ and $t_{flare}$).
To provide both the domain flux conservation and the increasing field non-potentiality, the MCC model includes currents only on the intersections between the domain boundaries,
called {\it separators}. In this way the lack of reconnection leads to storage of free magnetic energy, energy above that of the potential field, which could then be released
by reconnection in the flare.  To achieve the maximum energy release, the field inside the domains associated with the flare ({\it flaring domains}) would need to relax to
its potential state.  Thus to find the reconnection flux we first need to find the flaring domains and then calculate the changes in their domain flux from $t_0$ to $t_{flare}$. More specifically, we first overlay the magnetic point
charges rotated to the time of the TRACE 1600 \AA{} flare ribbons onto the ribbons image and then use a Monte
Carlo method \cite{Barnes2005} to find the fluxes of the
flaring domains at $t_0$ and $t_{flare}$.   Finally, we separately sum up the absolute values of all the positive and negative changes in the domain fluxes to calculate the model reconnection flux ($\Phi_{r,MCC}$, see Table~\ref{tab:t02_4}).
This is the model estimate of the net flux transfer that must occur in the two-ribbon flare through the flare reconnection.

To find the  {\it flaring separators} we find the topology of the magnetic field at $t_{flare}$ and select those separators that connect nulls that are located on the flare ribbons.
Through the MCC model, the changes in the domain fluxes under those flaring separators allow us to calculate current, free energy and helicity liberated on each
separator (for a detailed description of the method see \inlinecite{Longcope1996d} and Appendix B of \inlinecite{Kazachenko2009}). The total model energy
($\mathcal{E}_{MCC}$) released during the flare is a sum of energies released at each flaring separator. It is a lower bound on the energy stored by the pre-flare motions,
 since MCC model uses the point charge representation and hence applies a smaller number of constraints than point-for-point line-tying. It can be shown that the energy stored by ideal,
 line-tied, quasi-static evolution will always exceed the energy of the corresponding flux constrained equilibria \cite{Longcope2004}.
The total mutual helicity injected on all flaring separators is a sum of the helicities injected on each flaring separator. However the liftoff of the flux rope does not remove all of the helicity available in the flux rope \cite{Mackay2006}. For an MHD-simulated eruption \inlinecite{Gibson2008} found that $41\%$ of the helicity is lost with the escaping rope,  while $59\%$ remains. For simplicity, we assume that $50\%$ of the total mutual helicity from the MCC model ends up as self helicity of the flux rope created by reconnection:  $H_{MCC}=\sum H_{i}/{2}$.
Finally we note that the MCC model depends on the way we partition the magnetogram sequence \cite{Beveridge2006}. We experimented with different values of saddle points in the partitioning and apodizing windows in the LCT and found that this contributes an uncertainty in the MCC reconnection flux, MCC energy and MCC helicity that we include in Table~\ref{tab:t02_4}.

\subsection{Flare And MC Observations}\label{obs4}
For comparison to the predictions of the MCC model, we must infer values of reconnection flux $\Phi_{r,ribbon}$ from the observations, MC poloidal flux $\Phi_{p,MC}$, energy $\mathcal{E}_{GOES}$ and helicity $H_{MC}$ (see Table~\ref{tab:t02_4}).

To infer values of reconnection flux $\Phi_{r,ribbon}$ from the observations we use flare ribbon motion \cite{Poletto1986,Fletcher2001b} observed in 1600 \AA{} images from TRACE.
To find the total magnetic flux swept out by a moving ribbon, we count all pixels that brightened during any period of the flare and then integrate the unsigned magnetic flux
encompassed by the entire area taking into account the height of the ribbon's formation, a $\approx20\%$ correction \cite{Qiu2007}.  The uncertainties in the $\Phi_{r,ribbon}$ are
estimated by artificial misalignment between the MDI and TRACE data, ribbon edge uncertainty and inclusion of transient non-ribbon features with the ribbon areas.  To quantify the misalignment
 contribution we perform a set of trials whereby magnetogram and 1600 \AA{} images are offset by up to 2 MDI pixels. To find the uncertainty due to ribbon edge identification
we perform the calculation for different ribbon-edge cutoff values ranging from 6 to 10 times the background intensity. We also compare the MCC reconnection flux $\Phi_{r,MCC}$
to the poloidal MC flux $\Phi_{p,MC}$ derived from fits to the in-situ MC ACE/Wind observations using the Grad-Shafranov
reconstruction method
 \cite{Hu2001}. As \inlinecite{Qiu2007} showed from observations, $\Phi_{r,ribbon}\sim\Phi_{p,MC}$. Hence if the MCC model captures the reconnection flux correctly,
 $\Phi_{r,MCC}$ should match $\Phi_{p,MC}$ unless reconnection of the ICME with the ambient solar wind makes an important contribution \cite{Dasso2006}.

During the flare the magnetic free energy that has been slowly stored by photospheric motions $\mathcal{E}_{MCC}$ is rapidly released by reconnection
and then dissipated.  We estimate energy losses not only due to radiation  ($\mathcal{E}_{r}$),
 as \inlinecite{Kazachenko2009,Kazachenko2010} did, but also due to conductive cooling ($\mathcal{E}_{c}$) and the enthalpy flux ($\mathcal{E}_{ent}$), which in some numerical cases is as large as radiative energy losses \cite{Bradshaw2010}. Since it is not clear whether the source for the CME kinetic energy is the magnetic free energy stored in the active region and not the energy stored e.g. in the interplanetary current sheet, we neglect the energy carried away by the CME.

\begin{table*}
\caption{\small Observed energy budget (in $10^{31}$ ergs): radiative losses ($\mathcal{E}_{r}$), conductive losses ($\mathcal{E}_{c}$),
enthalpy fluxes ($\mathcal{E}_{ent}$), total energy ($\mathcal{E}_{GOES}=\mathcal{E}_{c}+\mathcal{E}_{r}+\mathcal{E}_{ent}$ and
estimated value for flare luminosity $\mathcal{E}_{FL}\approx(3.15\pm1.05)\times\mathcal{E}_{GOES}$. The predicted model energy ($\mathcal{E}_{MCC}$) is given for comparison with the observations (see \S\ref{obs4} and \S\ref{flares4}).} \begin{center}
\begin{tabular}{c|cccc|cc|c}
$i $ & $L$ (Mm) & $\mathcal{E}_{r}$ & $\mathcal{E}_{c}$ & $\mathcal{E}_{ent}$ & $\mathcal{E}_{GOES}$ & $\mathcal{E}_{FL}$ & $\mathcal{E}_{MCC}$ \\
\hline
1 & $145\pm31$ &1.0  & $0.45\pm0.05$ &   $1.6\pm0.6$  &  $3.1\pm0.6$ & $10.3\pm5.1$ & $1.0\pm0.3$\\
2 & $43\pm20$  &0.9  & $0.4\pm0.1$   &   $0.7\pm0.1$  &  $2.0\pm0.1$ & $6.4\pm2.4$ & $6.4\pm1.8$\\
3 & $151\pm50$ &2.5  & $2.6\pm0.2$   &   $4.9\pm1.9$  & $10.1\pm2.1$ & $34.0\pm17.2$ & $9.1\pm2.6$\\
4 & $107\pm18$ &5.3  & $2.75\pm1.35$ &   $5.5\pm0.7$  &  $13.6\pm0.6$ & $43.5\pm16.1$ & $18.0\pm5.2$\\
\end{tabular}
\normalsize \label{tab:t03_4} \end{center} \end{table*}
To quantify the three components of $\mathcal{E}_{GOES}=\mathcal{E}_{r}+\mathcal{E}_{c}+\mathcal{E}_{ent}$ (see Table~\ref{tab:t03_4}) we use GOES analysis software in SolarSoft and the observed GOES X-ray fluxes in the two channels (1--8 \AA{} and 0.5--4 \AA{}). Those  provide an estimate of the plasma temperature $T$ and emission measure $EM=n_e^{2}V$,  where $n_e$ is the electron density and V is the emitting volume. Radiative energy losses $\mathcal{E}_{r}$ depend on the emission measure, temperature  and composition of emitting plasma. We find their magnitude using the temperature dependent Mewe radiative loss function \cite{Mewe1985}.  To calculate the conductive energy losses $\mathcal{E}_{c}$ we integrate  the conductive energy loss rate to the
chromosphere $P_{cond}=U_{th}/\tau_{cond}$ where $U_{th}$ is thermal energy content of the plasma $U_{th}=3n_ekTV=3kT\sqrt{EM\times V}$ and $\tau_{cond}$ is the cooling time scale
\be
\tau_{cond}\simeq\frac{3kn_e(L/2)^2}{\kappa_0T^{5/2}},
\label{eq:tau}
\ee
for a loop of full length $L$, Boltzmann constant $k$ and Spitzer conductivity $\kappa_0\simeq10^{-6}$ \cite{Longcope2010}. We quantify the volume of the emitting material $V$ by assuming
that $\mathcal{E}_{c}\approx\mathcal{E}_{r}$ at late times, as they should be in a static equilibrium \cite{Rosner1978,Vesecky1979}. From the volume $V$ and emission measure $EM$ we
derive the electron density $n_e=\sqrt{EM/V}$. For the loop length $L$ we use the distribution of the lengths of the flaring separators (with the energy weights) whose geometrical properties
are found from the coronal magnetic topology at $t_{flare}$ (\S\ref{MCC4}). The mean and standard deviation of the lengths of the flaring separators
yield the mean and the standard deviation of the values of $\mathcal{E}_{c}$.
Finally, we estimate the enthalpy flux $\mathcal{E}_{ent}$ using model calculations by \inlinecite{Bradshaw2010}. From Tables 1 and 2 in the paper by \inlinecite{Bradshaw2010}
and the loop lengths of the flaring separators, we first derive a coefficient $\delta$ which describes the ratio between the radiative cooling and the enthalpy flux
 time scales and then the enthalpy flux itself.

We get an additional idea for the value of the uncertainty in $\mathcal{E}_{GOES}$ by comparing it to the flare luminosity (FL, $\mathcal{E}_{FL}$) from the Total Irradiance Monitor (TIM) on the Solar Radiation and Climate Experiment (SORCE). Unfortunately, FL measurements are not available  for the four flares studied here.  However FLs have been measured for four other large  ($>$X10) solar flares (see Table 2  in \inlinecite{Woods2006}), for which we may calculate $\mathcal{E}_{GOES}$.  For these four flares we find that the FLs are approximately two to four times larger than the $\mathcal{E}_{GOES}$. We use this scaling range to limit our energy estimates from above (see Table~\ref{tab:t03_4}): $\mathcal{E}_{FL}\approx(3.15\pm1.05)\times\mathcal{E}_{GOES}$.

Finally, we compare the model MCC flux rope helicity with the helicity of the magnetic cloud associated with the flare, $H_{MC}$. We calculate $H_{MC}$ applying the
Grad-Shafranov method \cite{Hu2001} to the ACE/Wind MC observations.
There are several uncertainties and limitations in the determination of $H_{MC}$, which are as well applicable to MC poloidal flux calculations.
First, the inferred value of $H_{MC}$ is model-dependent:
e.g. within the cylindrical hypothesis, force-free and non-force-free
models give helicities values that differ by up to $30\%$ \cite{Dasso2003,Dasso2006}. However, this variation remains small compared to the variation of helicity
values computed for different MCs \cite{Gulisano2005}. Second, the MC boundaries can be defined by several criteria,
which do not always agree. This introduces an uncertainty on the magnetic helicity which can be comparable to the uncertainty
obtained with different models \cite{Dasso2006}. Finally,
the distribution of the twist along the flux rope, as well as the length of the flux rope are generally not known.
So far only in one case the length of the flux rope $L_{MC}=2.5AU$ has been determined precisely from impulsive electron events and solar type III radio
bursts \cite{Larson1997}.  We take the value of $0.5$ AU as the lower limit of $L_{MC}$ \cite{DeVore2000_rcc} and $2.5$ AU as the upper limit of $L_{MC}$ \cite{Larson1997}. Such
choices of the lower and upper limits of $L_{MC}$ would change poloidal MC flux and MC helicity to vary between roughly half and twice the measured values.

\section{Data: Flares Studied}\label{flares4}
We apply the methods described in the previous section to four large eruptive flares (Table~\ref{tab:t01_4}).  This number of events is limited by several necessary flare selection criteria.
Firstly, we selected only events which have good observations of both the flare and the MC. Secondly, except for the May 13 2005 flare, we selected only ARs where two successive flares
 larger than M-class were present, in order to make plausible our assumption of initial relaxation of the AR's magnetic field to potential state. Thirdly, both the flare of study and
 the zero-flare should happen no farther than $40^{\circ}$ from the central meridian so that the stress build-up could be observed. Finally, we selected only flares associated with ARs with no significant flux emergence or cancellation
 during the period between $t_0$ and $t_{flare}$.

Our topological analysis using the MCC model has been executed previously for three of four flares: M8 flare on May 13 2005 (Kazachenko 2009), X2 flare on November 7 2004 \cite{Longcope2007},
 X17 flare on October 28 2003 \cite{Kazachenko2010}. The results of the MCC analysis for the X5.7 flare on July 14 2000 are described in this paper for the first time (see Appendix).
 In Table~\ref{tab:t01_4} we list the flare number ($i$) in this work, date, time and X-ray class of each flare; the NOAA number of the AR (AR) associated with the flare, AR's unsigned magnetic  flux ($\Phi_{AR}$) and helicity injected into the AR during the magnetogram sequence ($H_{AR}$); start ($t_{0}$) and end time ($t_{flare}$) of the magnetogram sequence. The flares are sorted by X-ray class. In Table~\ref{tab:t02_4} we compare MCC model predicted physical properties with the observations:
 predicted and inferred from the ribbon motions reconnection fluxes and MC poloidal fluxes, predicted and observed from the GOES observations energy values, predicted and observed from
 the Wind/ACE observations helicities. Finally in Table~\ref{tab:t03_4} we detail the observed energy budget for each flare.

The first flare listed in Table~\ref{tab:t01_4} is the M8 flare that occurred on May 13 2005 in NOAA 10759
\cite{Kazachenko2009,Yurchyshyn2006,Liu2007,Jing2007,Liu2008}. NOAA 10759 has a large positive sunspot which contains more than a half of the total positive flux of the AR and
 rotates with the rate of $0.85^{\circ}\pm0.13^{\circ}$ per hour during 40 hr before the flare \cite{Kazachenko2009}. Such fast rotation along with the fact that the spin helicity flux is proportional
 to the magnetic flux  squared makes the effect of sunspot rotation dominant in the helicity budget of the whole AR. As for the flare itself, the rotation of the sunspot
 produced three times more energy and
magnetic helicity than in the hypothetical case in which the sunspot does not rotate; the inclusion of sunspot rotation in the analysis brings the model into substantially better agreement
with GOES and interplanetary magnetic cloud observations. Rotation is energetically important in the flare and alone can store sufficient energy to power this M8 flare.

The second flare in Table~\ref{tab:t01_4} is the X2 flare on November 7 2004  \cite{Longcope2007}. The start time was plausibly taken to be that of an M9.3 flare which occurred 40 hr
before the flare of interest. The MCC model predicts a value of the flux needed to be reconnected in the flare that compares favorably with the flux swept up by the flare ribbons.
The MCC model places a lower bound on the energy stored by the 40-hour buildup shearing motions that is at least three times larger than the observed energy losses.
The helicity assigned to the flux rope that is assumed in the model is comparable to the magnetic cloud helicity.  Note that our estimate for  $H_{MCC}$ in Table~\ref{tab:t02_4} is higher than the one in  \inlinecite{Longcope2007} (see Table II in \inlinecite{Longcope2007}):  we estimate $H_{MCC}$ as a sum of the  helicities over all eight flaring separators, while \inlinecite{Longcope2007}
took a  sum over only the three most energetic separators.

The third flare in Table~\ref{tab:t01_4} is the X6 flare on July 14 2000 \cite{Lepping2001,Yurchyshyn2001,Fletcher2001b,Masuda2001,Bast2001}.
Our analysis of this flare is described in this paper for the first time (see Appendix). We take as the zero-flare an X1.9 flare around 48 hr before the flare of interest.
We use the MCC model to find that the released energy is comparable to the observed energy losses. The amount of flux reconnected during the flare according
 to the model is at least one and a half times smaller than the reconnection flux observed with TRACE. The model estimate for the helicity is comparable with the helicity from
 the MC observations. No sunspot rotation is associated with the pre-flare evolution.

The fourth event in Table~\ref{tab:t01_4}, the X17 Halloween flare \cite{Yurchyshyn2005,Regnier2007,Schrijver2006,Mandrini2006,Lynch2005,Zhang2008}, occurred in an AR
 with a fast-rotating sunspot. We find that the MCC reconnection flux is consistent with the reconnection flux inferred from the observations. We find that the sunspot rotation increases
 the total AR helicity by $\approx50\%$. However in contrast to the flare on May 13 2005, where rotation is dominant in the energetics, rotation increases the free energy and
 flux rope helicity of this flare by only $\approx10\%$. Shearing motions alone store sufficient energy and helicity to account for the flare energetics and ICME helicity content
within their observational uncertainties.
Thus this flare demonstrates that the relative importance of shearing and rotation in this flare depends critically on their location within the parent AR topology.

\section{Results: MCC Model Predictions Versus Observations}\label{discussion4}
\begin{figure*}
\includegraphics[angle=0,scale=0.6]{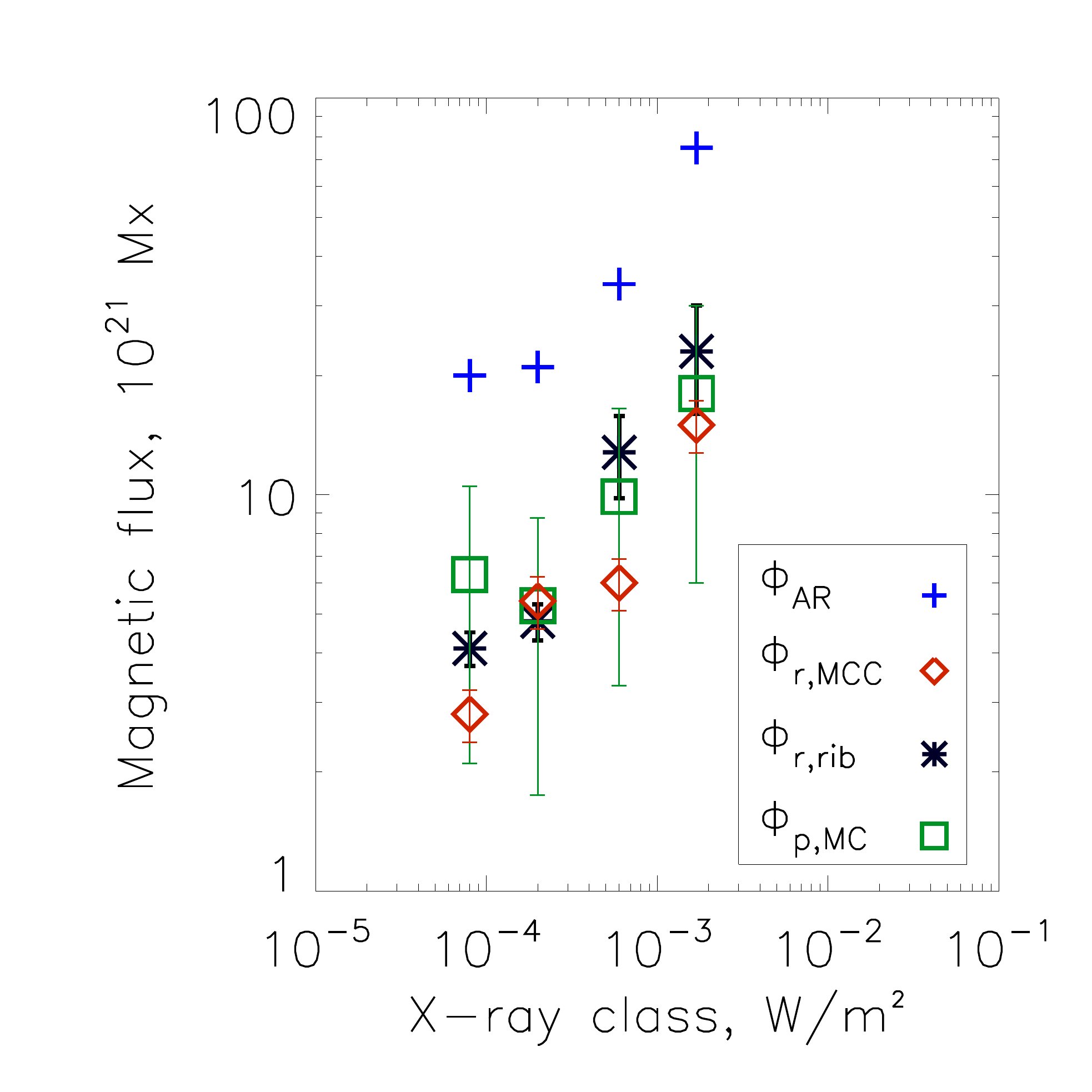}  
\caption{Predicted and observed magnetic flux values for the four events. The horizontal axis shows the maximum X-ray flux of each flare.
For details see Tables~\ref{tab:t01_4} and~\ref{tab:t02_4}. For discussion see \S\ref{discussion4}}.\label{fig-flux-f}
\end{figure*}

The main global property that describes the flare's reconnection is the amount of magnetic flux that participates. Figure~\ref{fig-flux-f} shows predicted ($\Phi_{r,MCC}$) and observed
($\Phi_{r,ribbon}$) reconnection fluxes for each event, the AR average unsigned magnetic flux ($\Phi_{AR}$) and the poloidal MC flux ($\Phi_{p,MC}$).
The first thing we notice is that the fraction of the AR magnetic flux that is observed to reconnect during the four flares ranges from $18\%$ to $49\%$. Secondly, in the second and fourth flares the predicted reconnection flux matches the reconnection flux inferred from the observations, while in the first and third flares the highest probable value of the MCC reconnection flux is lower than the lowest probable value of the observed reconnection flux by $13\%$  and $29\%$ correspondingly.
The lower model reconnection flux is likely due to additional reconnections not accounted for in the model. That means that the MCC model captured a lower limit of the amount of magnetic flux that has reconnected in these flares and hence the lower limit on the amount of energy released. Finally, in all four cases the value of poloidal MC flux matches both the observed and model reconnection fluxes, although the uncertainties in $\Phi_{p,MC}$ due to the unknown MC length are quite large. According to the CSHKP model, on which the MCC model builds, reconnection contributes solely to the incremental poloidal component of the flux-rope flux. Therefore, the derived agreement between the poloidal MC flux and the reconnection fluxes means that the flux rope is formed in situ.

\begin{figure*}
\includegraphics[angle=0,scale=0.6]{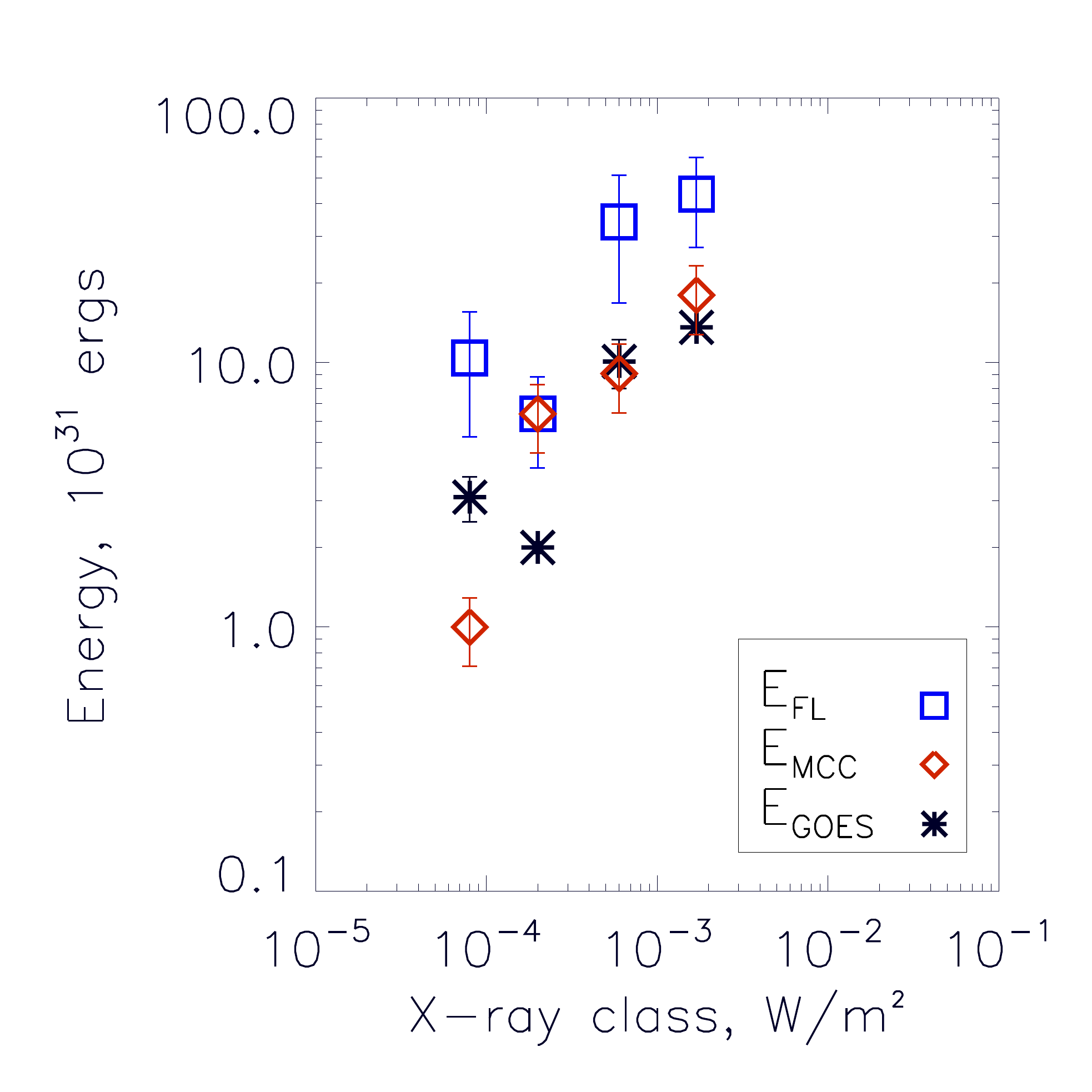}  
\caption{Predicted and observed energy values for the four events.  For details see Table~\ref{tab:t03_4}. For discussion see \S\ref{discussion4}}.  \label{fig:en_f}
\end{figure*}

The MCC model gives a lower limit of the free magnetic energy released in each flare \cite{Longcope2001b}. In Figure~\ref{fig:en_f} we compare the predicted MCC model free energy
($\mathcal{E}_{MCC}$, diamonds) with the observed time-integrated sum of radiative and conductive energy losses and the enthalpy flux ($\mathcal{E}_{GOES}$, stars); we also show the
estimated flare luminosity  ($\mathcal{E}_{FL}$, blue squares).  Figure~\ref{fig:en_f} indicates that for the third and fourth flares the predicted energy $\mathcal{E}_{MCC}$ matches the observed energy  $\mathcal{E}_{GOES}$, while for the second flare the predicted energy is around three times larger than $\mathcal{E}_{GOES}$. In all four cases the flare luminosity is higher than both $\mathcal{E}_{GOES}$ and $\mathcal{E}_{MCC}$. This is not surprising, since the MCC model uses a point charge representation rather than line tying and yields a lower limit on the free energy released in the flare. Summarizing, within the uncertainties, for three flares $i=2,3,4$ the predicted free energy lies between the observed estimate of released energy and the estimated flare luminosity.
Only for the May 13 2005 flare is the model energy lower than the observed estimate.  Note, that this flare is the only flare that did not have a zero-flare at $t_0$.
Since rotation is the dominant source of helicity injection in this flare  and the rotation rate was around zero before $t_0$, we believe that our analysis
plausibly captures the major source of helicity injection in this flare. Nevertheless, the absence of a zero-flare indicates that there might have been additional energy storage
before $t_0$. Hence, our estimate is a lower limit to the reconnection flux and the magnetic energy of this M8 flare.

One must keep in mind that one of the basic assumptions of the MCC model is the potentiality of the magnetic field after the zero-flare. \inlinecite{Su2007} analyzed TRACE observations of 50 X- and M-class two-ribbon flares and found that $86\%$ of these flares show a general decrease in the shear angle between the main polarity-inversion line and pairs of conjugate bright ribbon kernels.  They interpreted this as a relaxation of the field towards a more potential state because of the eruption that carries helicity/current with it, but one can readily argue that a similar decrease in shear angle would be seen if sequentially higher, less-sheared post-flare loops light up with time as the loops cool after reconnection.  These results are consequently ambiguous: they may show a decrease in shear, or they may reflect that flares generally do not release all available energy and part of the flux-rope configuration remains. In other words, the MCC model potentiality assumption may mean that additional energy and reconnection flux is stored before the zero-flare.

\begin{figure*}
\includegraphics[angle=0,scale=0.6]{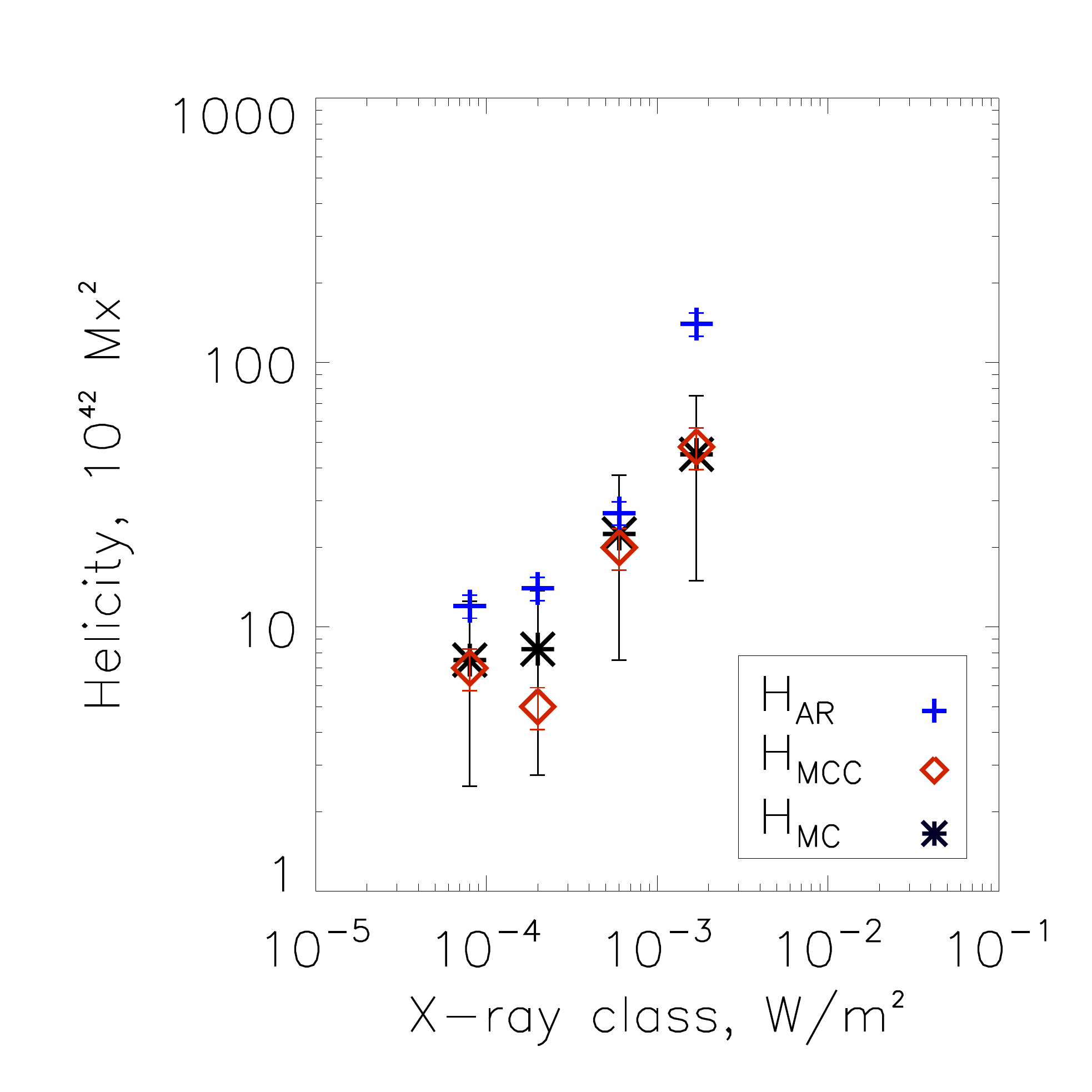}  
\caption{Predicted and observed helicity values for the four events. For details see Table~\ref{tab:t01_4} and Table~\ref{tab:t02_4}. For discussion see \S\ref{discussion4}.}
 \label{fig:hel_f}
\end{figure*}

Because magnetic helicity is approximately conserved in the corona, even in the presence of reconnection, it is instructive to compare $H_{MCC}$ to $H_{MC}$. Figure~\ref{fig:hel_f} shows the
relationship between the predicted ($H_{MCC}$, diamonds) and observed ($H_{MC}$, stars) values of helicity. The blue $+$s show the amount of the helicity for the whole AR
($H_{AR}$).
In all four cases we get MCC model helicities
that are of comparable magnitude and same sign as the observed MC values and are smaller than the helicity of the whole AR. Our analysis shows that preflare motions
contribute enough stress to account for observed helicity values, however more accurate estimate of the MC length is required in order to lower the error bars in the MC helicities and
improve our understanding of MC/flare relationship.
It is interesting that the most energetic of the four flares ($i=4$, which happened in the southern hemisphere) had the helicity sign opposite to the hemispheric helicity preference
\cite{Pevtsov2003d}. Hence its sign cannot be predicted from the global solar properties, but only from a case study like this.

The properties of the magnetic field in a MC are determined by the initial conditions of the eruption which we derive with the MCC model as well as by how the MC interacts
with the interplanetary medium during its travel toward the Earth. Above we found a consistency between the MC flux and predicted model and observed flare reconnection fluxes,
the MC and predicted flare flux rope helicities, observed and predicted flare energy releases. The agreement between those supports our local in-situ formed flux rope
hypothesis.

One more quantity that is frequently compared between the MC and AR flux rope is the direction of the poloidal field.
\inlinecite{Li2010} found that the poloidal field of MCs with low axis inclination relative to the ecliptic ($\approx40\%$ of all MCs) has a solar cycle dependence.
They note that during the solar minima, the orientation of the leading edge of
the MC is predictable: it is the same as the solar dipole field. However during the maximum and the declining phases, when most of the geoffective MCs happen,
both (north and south) orientations are present, although the global dipole field orientation of the beginning of the cycle dominates.

It is instructive to consider how the four flares of our study relate to these results. Assuming that the poloidal field in
the ejected flux rope is oriented along the flaring separators, we define its orientation relative to the ecliptic plane using the North-South classification: north ($30^{\circ}<\theta<90^{\circ}$, $B_z>0$, in the solar ecliptic coordinate system) or south
($-90^{\circ}<\theta<-30^{\circ}$, $B_z<0$).  We then determine the orientation of the leading edge of the MC poloidal field using the same North-South classification and compare two quantities:  the orientation of the poloidal field in the active region and the orientation of the leading edge of the MC poloidal field. We find that the MC produced by the Bastille day flare during the solar maximum has a south oriented leading MC poloidal field, same as both the remnant weak dipole orientation and poloidal field orientation of the flux rope at the sun. In contrast, the flares which occurred during the declining phase, on May 13 2005 and  November 7 2004, produced magnetic clouds with south oriented leading MC poloidal fields, opposite to the direction of the global dipole field, but same as the poloidal field orientation predicted for a flux rope in the modeled AR.
Finally the MC produced by the Halloween flare laid perpendicular to the ecliptic plane and thus was not relevant to the observed \inlinecite{Li2010} rule; however a good agreement was also found between the directions of the poloidal field in the MC and in the source AR \cite{Yurchyshyn2005}.

Summarizing the above, although there is a tendency for ARs to follow the dipole field orientation during the solar minimum \cite{Li2010}, during the solar maximum and the declining phase, when the largest MCs occur, the local AR field is important.
We find that for the four studied large events the direction of the leading MC poloidal field is consistent with the poloidal field orientation in the AR rather than to the global dipole field in agreement with \inlinecite{Leamon.Canfield.Pevtsov2002_rcc}.  This implies that the magnetic clouds associated with large ARs inherit the properties of the AR rather than those of the global dipole field, as a result of reconnection in the active region rather than with the surrounding dipole field. Although here we compare the poloidal post-flare arcade field with the poloidal MC field, this supports the conclusion by \inlinecite{Yurchyshyn2007}, who found that  $64\%$  of CMEs are oriented within $45^{\circ}$ to the MC axes (MC toroidal field)
and $70\%$ of CMEs are oriented within $10^{\circ}$ to the toroidal field of EUV post-flare arcades \cite{Yurchyshyn2009}. In other words, despite the fact that CME flux ropes may interact significantly with the ambient solar wind \cite{Dasso2006} or other flux ropes \cite{Gopalswamy2001}, a significant group of MCs reflects the magnetic field orientation of the source regions in the low corona.

\section{Conclusions}\label{conclusion}
The main purpose of this study is to understand the mechanism of the CME flux rope formation and its relationship with the MC.  Notably, we use the Minimum Current Corona model \cite{Longcope1996d} which, using the pre-flare motions of photospheric magnetic fields and flare ribbon observations, quantifies the
reconnection flux, energy and helicity budget of the flare. We apply this model to four major eruptive solar flares that produced MCs and compare the predicted flux rope properties to the observations.

We compare model predictions to observations of four quantities:
the predicted model reconnection fluxes to the MC poloidal fluxes and ribbon motion reconnection fluxes, the predicted flux rope helicities to the MC helicities, the predicted released energies to the total radiative/conductive energy losses plus the enthalpy fluxes, the direction of the magnetic field in the AR arcade to the direction of the leading edge of MC poloidal field.

Our comparison reveals the following. The predicted reconnection fluxes match the reconnection fluxes inferred from the observations for the November 7 2004 and Halloween flares. For the May 13 2005 and Bastille day flares the minimum probable differences between the predicted and observed reconnection fluxes are $13\%$ and $29\%$ correspondingly.
In all four cases the values of poloidal MC fluxes match both the observed and the model reconnection fluxes.
The predicted flux rope helicities match the MC helicities.
For three flares of study the predicted free energies lie between the observed energy losses (radiative and conductive energy losses plus the enthalpy fluxes) and the flare luminosities. Only for the flare on May 13 2005, the predicted free energy is one third of
the observed estimate. We relate this mismatch to the fact that May 13 2005 flare was the only event without a zero-flare, hence additional energy might have been stored before $t_0$.
Finally, we find that in all four cases the direction of the leading MC poloidal field is consistent with the poloidal component of the local AR arcade field, whereas in two cases the MC poloidal field orientation is opposite to that of the global solar dipole.

These findings compel us to believe that magnetic clouds associated with these four eruptive solar flares are formed by low-corona magnetic reconnection
during the eruption, rather than eruption of preexisting structures in the corona or formation in the upper corona by the global field. Our findings support the conclusions of \inlinecite{Qiu2007} and \inlinecite{Leamon2004}, although through a very different approach: while \inlinecite{Qiu2007} and \inlinecite{Leamon2004} inferred the solar flux rope properties only from observations, we infer them from both the MCC model and the observations. Using the pre-flare magnetic field evolution and the MCC model, we find that we are able to predict the observed reconnection fluxes within a $29\%$ uncertainty and the observed MC poloidal flux and helicity values within the MC length uncertainty. For the flares associated with zero-flares we are able to estimate a lower limit for the free magnetic energy.  We note that, since all four flares occurred in ARs without significant pre-flare flux emergence/cancellation, the flux/energy/helicity we find is stored by shearing and rotating motions, which is sufficient to account for observed energy and MC flux and helicity.

Our work brings up several interesting questions that require further exploration. Firstly, the results of this paper are based on only a small number of events, which are similar in that all have a large radiative signature. Hence a study of observations  of a class of events with small radiative signature would be challenging: smaller flares that are nevertheless associated with major CMEs \cite{Aschwanden2009}. If the MCC model is valid, it should be able to explain both the energy and helicity content of flare/CME events whose flare energy output is disproportionately small.  Secondly, in this paper, we estimated the final flare energy as a sum of energy losses by radiation, conduction and enthalpy, neglecting the energy carried away by the CME.  To our knowledge, no systematic empirical study of the source of energy for the CME has yet been conducted, so it is unknown what proportion of its energy budget arises from the AR.  From the limited sample, \inlinecite{Ravindra2010} found that a $50\%$ contribution may be a reasonable first-order approximation.  Including CME energy losses into our flare analysis would help us understand how much of the CME energy arises from the active region and may lead to a greater understanding of the onset mechanism for CMEs. 

\appendix \label{apc}

\begin{figure}
\advance\leftskip-0.2cm
\includegraphics
[angle=0,scale=0.65,bb=30 0 0 410]
{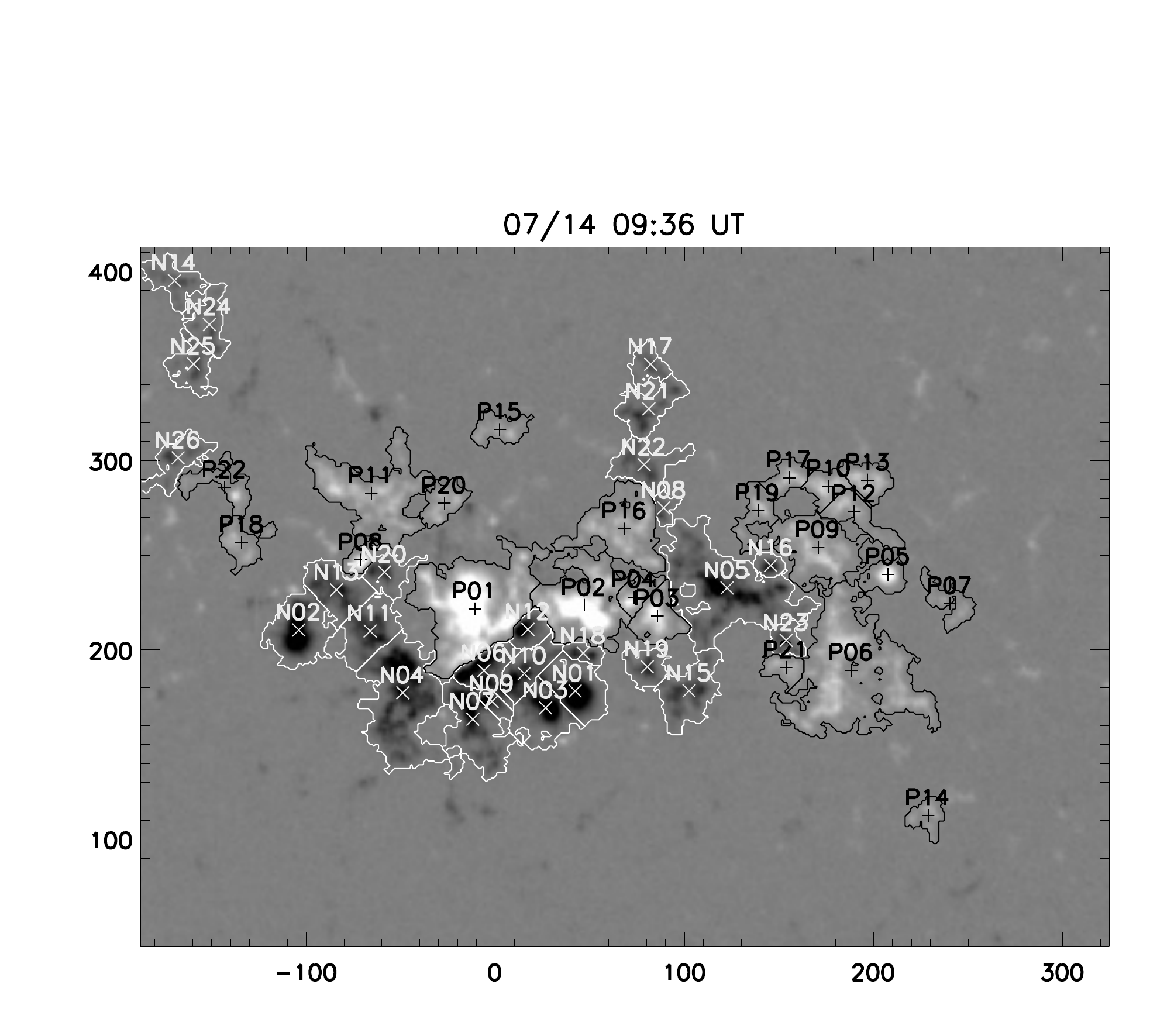} \caption{~Positive (P) and negative (N) polarity partitions for NOAA 9077 on July 14 09:36 UT. The gray-scale magnetogram shows magnetic
field scaled from -1000G to 1000G. The
partitions are outlined and the centroids are denoted by +'s and x's
(positive and negative respectively). Axes are labeled in arc-seconds from disk center.} \label{fig:f04_4}
\end{figure}

The X5.7 Bastille Day flare occurred on 2000 July 14 2000 at 10:03 UT in NOAA 9077.
 Our  magnetic field data describing the evolution of the magnetic field before this flare consist of a sequence of SOI/MDI full-disk magnetograms (2'', level 1.8) starting at $t_{0}=$ 2000
 July 12 14:27 UT, after the X1.9 flare (2000 July 12 10:18 UT), and ending at $t_{flare}=$July 14 09:36 UT, 27 minutes before the Bastille day X5.7 flare. Thus we form a
 sequence of 28 low-noise magnetograms with a 96-minute cadence, which cover 43 hours of the stress buildup prior to the X5.7 flare on 2000 July 14 10:03 UT.
 Firstly, for all successive pairs of magnetograms we use a Gaussian apodizing
 window of 7'' to derive a local correlation tracking (LCT) velocity. We then take a magnetogram at $t_{flare}$ and group pixels, exceeding a threshold $B_{thr}=45$ Gauss downhill from
 each local maximum, into individual partitions. We combine partitions by eliminating any boundary whose saddle point is less than $350$ Gauss below either maximum it separates.
  Each partition is assigned a unique label which maintains through the sequence by using the LCT velocity pattern.  Figure~\ref{fig:f04_4} shows the spatial distribution of these
  partitions at $t_{flare}$. For expediting the assessment of the
field's connectivity we represent each magnetic partition with a magnetic point charge which contains the magnetic flux of the whole partition concentrated in the partition's centroid.
We find that the magnetic field is well balanced at $t_{flare}$ ($\Phi_+(\Phi_-)=3.3(-3.5)\times10^{22}$Mx) and exhibits no significant emergence/cancellation during the 43 hours
of pre-flare stress buildup time. From the LCT velocity and magnetic field in each point we find the flux of relative helicity into the corona to be $H_{AR}=-(27\pm2.7)\times10^{42}$Mx$^2$,
 no significant spin helicity content (rotation) has been detected.

\begin{figure*}
\advance\leftskip-1.2cm
\includegraphics
[angle=90,scale=0.5,bb=20 0 448 760]
{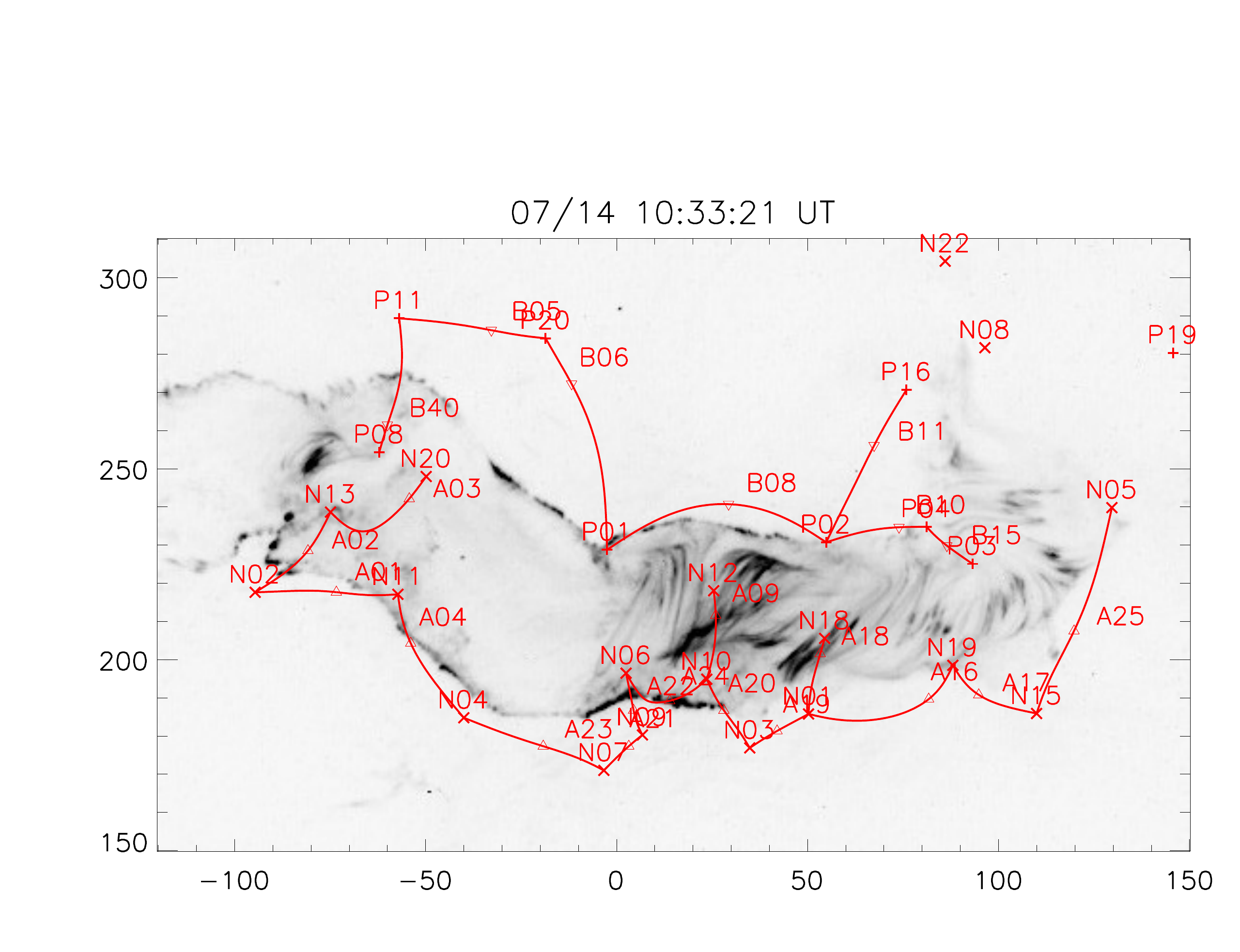}\caption{~TRACE 1600 \AA{}
image, plotted as reverse gray scale, with elements of the topological skeleton superimposed. The skeleton calculated for July 14 09:36 UT is projected onto the sky after its tangent
plane has been rotated to the time of the TRACE observations (10:33 UT). Positive and negative sources are indicated
by +'s and x's respectively. The triangles represent the labeled
null points. The red curved line segments show spine lines associated
with the reconnecting domains. Axes are in arc-seconds from disk center.}
\label{fig:f05_4}
\end{figure*}

To find the model estimate of the reconnection flux we determine the magnetic point charges associated with the flare using the flare UV observations by TRACE 1600 \AA{}.  Figure~\ref{fig:f05_4}
shows a superposition of the elements of the topological skeleton at $t_{flare}$ onto the UV flare ribbon image. The spines (red solid lines) that are associated with ribbons form the footprint
 of a combination of separatrices which overlay the flaring domains. The overlay suggests that the northern ribbon is associated with the spines connecting flaring point sources P08, P11, P20,
  P01, P02, P16, P04, P03; and the southern ribbon is associated with the spines connecting N20, N13, N02, N11, N04, N07, N06, N09, N10, N12, N03, N01, N18, N19, N15, N05. Field lines connecting
   the pairs of opposite point charges listed above form a set of flaring domains. The amount of flux that those domains exchanged, the model reconnection flux, is
   $\Phi_{r,MCC}=(6.0\pm0.9)\times10^{21}$Mx, fifty percent smaller than the lower value of the observed reconnection flux from the flare ribbon evolution ($\Phi_{r,ribbon}=12.8\pm3\times10^{21}$Mx).

\begin{figure*}
\advance\leftskip-1.2cm
\includegraphics
[angle=0,scale=0.75]
{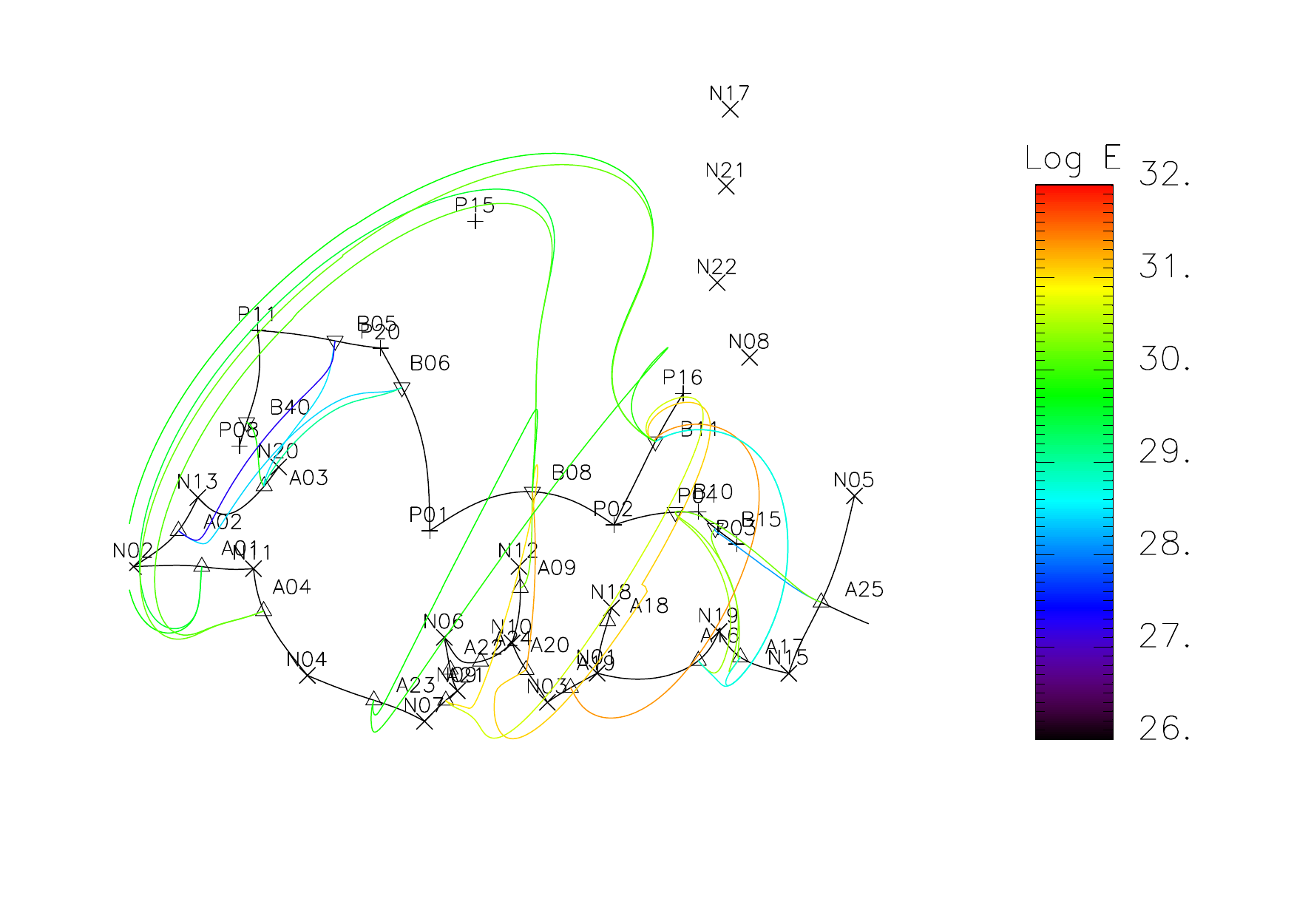}\caption{~Flaring separators derived from the MCC model. Colors indicate the logarithm of the free energy
(in ergs) available for release during the flare on each separator.}
\label{fig:f06_4}
\end{figure*}

From the set of flaring point charges and nulls lying between them we find twenty six flaring separators (see Figure~\ref{fig:f06_4}). The total free energy and helicity output on those
separators is $\mathcal{E}_{MCC}=(9.1\pm2.6)\times10^{31}$ergs and $H_{MCC}=-(20.1\pm3.6)\times10^{42}$Mx$^{2}$. However, out of the 26 flaring separators $90\%$ of the total free energy is contained
 in separators originating in nulls B08 and B11 which lie between poles P01, P02 and P16 (note the most red separators in Figure~\ref{fig:f06_4}). Moreover, $64\%$ of the total flare free
 energy is partitioned between three separators: A19/B11, A20/B11, A20/B08. According to the MCC model the poles associated with those nulls (P01, P02, P16 and N10, N01, N03, N18)
  indicate the locations of the largest free energy release. Figure~\ref{fig:f06_4} indicates that the brightest observed loops in TRACE 1600 \AA{} are the loops connecting point charges
  N12, N10, N01 and N18 with P02, consistent with the results of the MCC model presented above.

\begin{acks}
We thank the TRACE and SOHO MDI teams for providing the data. SOHO is a project of international cooperation between ESA and NASA. We are pleased to acknowledge support from NASA Earth and Space Science Fellowship grant NNX07AU73H (MDK), NASA LWS TR\&T grant NNG05-GJ96G (RCC and MDK).
\end{acks}

\bibliographystyle{spr-mp-sola}


\begin{thebibliography}{84}
\ifx \bisbn   \undefined \def \bisbn  #1{ISBN #1}\fi
\ifx \binits  \undefined \def \binits#1{#1}\fi
\ifx \bauthor  \undefined \def \bauthor#1{#1}\fi
\ifx \batitle  \undefined \def \batitle#1{#1}\fi
\ifx \bjtitle  \undefined \def \bjtitle#1{\textit{#1}}\fi
\ifx \bvolume  \undefined \def \bvolume#1{\textbf{#1}}\fi
\ifx \byear  \undefined \def \byear#1{#1}\fi
\ifx \bissue  \undefined \def \bissue#1{#1}\fi
\ifx \bfpage  \undefined \def \bfpage#1{#1}\fi
\ifx \blpage  \undefined \def \blpage #1{#1}\fi
\ifx \burl  \undefined \def \burl#1{\textsf{#1}}\fi
\ifx \href  \undefined \def \href#1#2{\textsf{#2}}\fi
\ifx \doiurl  \undefined \def
  \doiurl#1{\href{http://dx.doi.org/#1}{\textsf{#1}}}\fi
\ifx \betal  \undefined \def \betal{\textit{et al.}}\fi
\ifx \binstitute  \undefined \def \binstitute#1{#1}\fi
\ifx \bctitle  \undefined \def \bctitle#1{#1}\fi
\ifx \beditor  \undefined \def \beditor#1{#1}\fi
\ifx \bpublisher  \undefined \def \bpublisher#1{#1}\fi
\ifx \bbtitle  \undefined \def \bbtitle#1{\textit{#1}}\fi
\ifx \bedition  \undefined \def \bedition#1{#1}\fi
\ifx \bseriesno  \undefined \def \bseriesno#1{\textbf{#1}}\fi
\ifx \blocation  \undefined \def \blocation#1{#1}\fi
\ifx \bsertitle  \undefined \def \bsertitle#1{\textit{#1}}\fi
\ifx \bsnm \undefined \def \bsnm#1{#1}\fi
\ifx \bsuffix \undefined \def \bsuffix#1{#1}\fi
\ifx \bparticle \undefined \def \bparticle#1{#1}\fi
\ifx \barticle \undefined \def \barticle#1{}\fi
\ifx \botherref \undefined \def \botherref#1{}\fi
\ifx \url \undefined \def \url#1{\textsf{#1}}\fi
\ifx \bchapter \undefined \def \bchapter#1{}\fi
\ifx \bbook \undefined \def \bbook#1{}\fi
\ifx \bcomment \undefined \def \bcomment#1{#1}\fi
\ifx \oauthor \undefined \def \oauthor#1{#1}\fi
\ifx \citeauthoryear \undefined \def \citeauthoryear#1{#1}\fi
\def \endbibitem {}

\bibitem[\protect\citeauthoryear{{Abbett} and {Fisher}}{2003}]{Abbett2003}
\begin{barticle}
\bauthor{\bsnm{{Abbett}}, \binits{W.P.}}, \bauthor{\bsnm{{Fisher}},
  \binits{G.H.}}:
\byear{2003},
\batitle{A coupled model for the emergence of active region magnetic flux into
  the solar corona}.
\bjtitle{\apjl}
\bvolume{582},
\bfpage{475}\,--\,\blpage{485}.
doi:\doiurl{10.1086/344613}.
\end{barticle}
\endbibitem

\bibitem[\protect\citeauthoryear{Antiochos, Devore, and
  Klimchuk}{1999}]{Antiochos1999}
\begin{barticle}
\bauthor{\bsnm{Antiochos}, \binits{S.K.}}, \bauthor{\bsnm{Devore},
  \binits{C.R.}}, \bauthor{\bsnm{Klimchuk}, \binits{J.A.}}:
\byear{1999},
\batitle{A model for solar coronal mass ejections}.
\bjtitle{\apjl}
\bvolume{510},
\bfpage{485}\,--\,\blpage{493}.
\end{barticle}
\endbibitem

\bibitem[\protect\citeauthoryear{{Aschwanden},{Wuelser}, and
{Nitta}}{2009}]{Aschwanden2009}
\begin{barticle}
\bauthor{\bsnm{Aschwanden}, \binits{M.J.}}, \bauthor{\bsnm{Wuelser},
  \binits{J.P.}}, \bauthor{\bsnm{Nitta}, \binits{N.V.}}:
\byear{2009},
\batitle{Solar flares, Coronal mass ejections, EUV, Stereoscopy}.
\bjtitle{\solphys}
\bvolume{256},
\bfpage{3}\,--\,\blpage{40}.
\end{barticle}
\endbibitem

\bibitem[\protect\citeauthoryear{{Barnes}, {Longcope}, and
  {Leka}}{2005}]{Barnes2005}
\begin{barticle}
\bauthor{\bsnm{{Barnes}}, \binits{G.}}, \bauthor{\bsnm{{Longcope}},
  \binits{D.W.}}, \bauthor{\bsnm{{Leka}}, \binits{K.D.}}:
\byear{2005},
\batitle{{Implementing a Magnetic Charge Topology Model for Solar Active
  Regions}}.
\bjtitle{\apj}
\bvolume{629},
\bfpage{561}\,--\,\blpage{571}.
doi:\doiurl{10.1086/431175}.
\end{barticle}
\endbibitem

\bibitem[\protect\citeauthoryear{Berger}{1999}]{Berger1999}
\begin{bchapter}
\bauthor{\bsnm{Berger}, \binits{M.A.}}:
\byear{1999},
\bctitle{Magnetic helicity in space physics}.
In: \beditor{\bsnm{Brown}, \binits{M.R.}}, \beditor{\bsnm{Canfield},
  \binits{R.C.}}, \beditor{\bsnm{Pevtsov}, \binits{A.A.}} (eds.)
\bbtitle{Magnetic Helicity in Space and Laboratory Plasmas},
\bsertitle{Geophysical Monograph}
\bseriesno{111},
\bpublisher{AGU Press},
\blocation{Washington, DC},
\bfpage{1}\,--\,\blpage{9}.
\end{bchapter}
\endbibitem

\bibitem[\protect\citeauthoryear{{Berger} and {Field}}{1984}]{Berger1984}
\begin{barticle}
\bauthor{\bsnm{{Berger}}, \binits{M.A.}}, \bauthor{\bsnm{{Field}},
  \binits{G.B.}}:
\byear{1984},
\batitle{{The topological properties of magnetic helicity}}.
\bjtitle{Journal of Fluid Mechanics}
\bvolume{147},
\bfpage{133}\,--\,\blpage{148}.
doi:\doiurl{10.1017/S0022112084002019}.
\end{barticle}
\endbibitem

\bibitem[\protect\citeauthoryear{Beveridge and Longcope}{2006}]{Beveridge2006}
\begin{barticle}
\bauthor{\bsnm{Beveridge}, \binits{C.}}, \bauthor{\bsnm{Longcope},
  \binits{D.W.}}:
\byear{2006},
\batitle{A hierarchical application of the minimum current corona}.
\bjtitle{\apjl}
\bvolume{636},
\bfpage{453}\,--\,\blpage{461}.
\end{barticle}
\endbibitem

\bibitem[\protect\citeauthoryear{{Bothmer} and {Schwenn}}{1998}]{Bothmer1998}
\begin{barticle}
\bauthor{\bsnm{{Bothmer}}, \binits{V.}}, \bauthor{\bsnm{{Schwenn}},
  \binits{R.}}:
\byear{1998},
\batitle{The structure and origin of magnetic clouds in the solar wind}.
\bjtitle{\anngeo}
\bvolume{16},
\bfpage{1}\,--\,\blpage{24}.
\end{barticle}
\endbibitem

\bibitem[\protect\citeauthoryear{{Bradshaw} and {Cargill}}{2010}]{Bradshaw2010}
\begin{barticle}
\bauthor{\bsnm{{Bradshaw}}, \binits{S.J.}}, \bauthor{\bsnm{{Cargill}},
  \binits{P.J.}}:
\byear{2010},
\batitle{{The Cooling of Coronal Plasmas. III. Enthalpy Transfer as a Mechanism
  for Energy Loss}}.
\bjtitle{\apj}
\bvolume{717},
\bfpage{163}\,--\,\blpage{174}.
doi:\doiurl{10.1088/0004-637X/717/1/163}.
\end{barticle}
\endbibitem

\bibitem[\protect\citeauthoryear{{Burlaga} \textit{et~al.}}{1981}]{Burlaga1981}
\begin{barticle}
\bauthor{\bsnm{{Burlaga}}, \binits{L.}}, \bauthor{\bsnm{{Sittler}},
  \binits{E.}}, \bauthor{\bsnm{{Mariani}}, \binits{F.}},
  \bauthor{\bsnm{{Schwenn}}, \binits{R.}}:
\byear{1981},
\batitle{Magnetic loop behind an interplanetary shock - voyager, helios, and
  imp 8 observations}.
\bjtitle{\jgr}
\bvolume{86},
\bfpage{6673}\,--\,\blpage{6684}.
\end{barticle}
\endbibitem

\bibitem[\protect\citeauthoryear{{Carmichael}}{1964}]{carmichael1964}
\begin{bchapter}
\bauthor{\bsnm{{Carmichael}}, \binits{H.}}:
\byear{1964},
\bctitle{A process for flares}.
In: \beditor{\bsnm{Hess}, \binits{W.N.}} (ed.)
\bbtitle{AAS-NASA Symposium on the Physics of Solar Flares},
\bpublisher{NASA},
\blocation{Washington, DC},
\bfpage{451}.
\end{bchapter}
\endbibitem

\bibitem[\protect\citeauthoryear{{Chae}}{2001}]{Chae2001}
\begin{barticle}
\bauthor{\bsnm{{Chae}}, \binits{J.}}:
\byear{2001},
\batitle{{Observational Determination of the Rate of Magnetic Helicity
  Transport through the Solar Surface via the Horizontal Motion of Field Line
  Footpoints}}.
\bjtitle{\apjl}
\bvolume{560},
\bfpage{L95}\,--\,\blpage{L98}.
doi:\doiurl{10.1086/324173}.
\end{barticle}
\endbibitem

\bibitem[\protect\citeauthoryear{{Chae}}{2007}]{Chae2007}
\begin{barticle}
\bauthor{\bsnm{{Chae}}, \binits{J.}}:
\byear{2007},
\batitle{{Measurements of magnetic helicity injected through the solar
  photosphere}}.
\bjtitle{Advances in Space Research}
\bvolume{39},
\bfpage{1700}\,--\,\blpage{1705}.
doi:\doiurl{10.1016/j.asr.2007.01.035}.
\end{barticle}
\endbibitem

\bibitem[\protect\citeauthoryear{Chen}{1989}]{Chen1989}
\begin{barticle}
\bauthor{\bsnm{Chen}, \binits{J.}}:
\byear{1989},
\batitle{Effects of toroidal forces in current loops embedded in a background
  plasma}.
\bjtitle{\apjl}
\bvolume{338},
\bfpage{453}\,--\,\blpage{470}.
\end{barticle}
\endbibitem

\bibitem[\protect\citeauthoryear{{Crooker}}{2000}]{Crooker2000_rcc}
\begin{barticle}
\bauthor{\bsnm{{Crooker}}, \binits{N.U.}}:
\byear{2000},
\batitle{{Solar and Heliospheric Geoeffective Disturbances}}.
\bjtitle{JASTP}
\bvolume{62},
\bfpage{1071}\,--\,\blpage{1085}.
\end{barticle}
\endbibitem

\bibitem[\protect\citeauthoryear{{Dasso} \textit{et~al.}}{2003}]{Dasso2003}
\begin{barticle}
\bauthor{\bsnm{{Dasso}}, \binits{S.}}, \bauthor{\bsnm{{Mandrini}},
  \binits{C.H.}}, \bauthor{\bsnm{{D{\'e}moulin}}, \binits{P.}},
  \bauthor{\bsnm{{Farrugia}}, \binits{C.J.}}:
\byear{2003},
\batitle{Magnetic helicity analysis of an interplanetary twisted flux tube}.
\bjtitle{\jgr}
\bvolume{108},
\bfpage{3}\,--\,\blpage{1}.
doi:\doiurl{10.1029/2003JA009942}.
\end{barticle}
\endbibitem

\bibitem[\protect\citeauthoryear{{Dasso} \textit{et~al.}}{2006}]{Dasso2006}
\begin{barticle}
\bauthor{\bsnm{{Dasso}}, \binits{S.}}, \bauthor{\bsnm{{Mandrini}},
  \binits{C.H.}}, \bauthor{\bsnm{{D{\'e}moulin}}, \binits{P.}},
  \bauthor{\bsnm{{Luoni}}, \binits{M.L.}}:
\byear{2006},
\batitle{{A new model-independent method to compute magnetic helicity in
  magnetic clouds}}.
\bjtitle{\aap}
\bvolume{455},
\bfpage{349}\,--\,\blpage{359}.
doi:\doiurl{10.1051/0004-6361:20064806}.
\end{barticle}
\endbibitem

\bibitem[\protect\citeauthoryear{{D{\'e}moulin}}{2008}]{Demoulin2008}
\begin{barticle}
\bauthor{\bsnm{{D{\'e}moulin}}, \binits{P.}}:
\byear{2008},
\batitle{{A review of the quantitative links between CMEs and magnetic
  clouds}}.
\bjtitle{\anngeo}
\bvolume{26},
\bfpage{3113}\,--\,\blpage{3125}.
\end{barticle}
\endbibitem

\bibitem[\protect\citeauthoryear{{D{\'e}moulin} and
  {Pariat}}{2007}]{Demoulin2007}
\begin{barticle}
\bauthor{\bsnm{{D{\'e}moulin}}, \binits{P.}}, \bauthor{\bsnm{{Pariat}},
  \binits{E.}}:
\byear{2007},
\batitle{Computing magnetic energy and helicity fluxes from series of
  magnetograms}.
\bjtitle{\memit}
\bvolume{78},
\bfpage{136}.
\end{barticle}
\endbibitem

\bibitem[\protect\citeauthoryear{{D{\'e}moulin}
  \textit{et~al.}}{2002}]{Demoulin2002}
\begin{barticle}
\bauthor{\bsnm{{D{\'e}moulin}}, \binits{P.}}, \bauthor{\bsnm{{Mandrini}},
  \binits{C.H.}}, \bauthor{\bsnm{{van Driel-Gesztelyi}}, \binits{L.}},
  \bauthor{\bsnm{{Thompson}}, \binits{B.J.}}, \bauthor{\bsnm{{Plunkett}},
  \binits{S.}}, \bauthor{\bsnm{{Kov{\'a}ri}}, \binits{Z.}},
  \bauthor{\bsnm{{Aulanier}}, \binits{G.}}, \bauthor{\bsnm{{Young}},
  \binits{A.}}:
\byear{2002},
\batitle{What is the source of the magnetic helicity shed by cmes? the
  long-term helicity budget of ar 7978}.
\bjtitle{\aap}
\bvolume{382},
\bfpage{650}\,--\,\blpage{665}.
doi:\doiurl{10.1051/0004-6361:20011634}.
\end{barticle}
\endbibitem

\bibitem[\protect\citeauthoryear{{DeVore}}{2000}]{DeVore2000_rcc}
\begin{barticle}
\bauthor{\bsnm{{DeVore}}, \binits{C.R.}}:
\byear{2000},
\batitle{{Magnetic Helicity Generation by Solar Differential Rotation}}.
\bjtitle{\apjl}
\bvolume{539},
\bfpage{944}\,--\,\blpage{953}.
\end{barticle}
\endbibitem

\bibitem[\protect\citeauthoryear{{Fan} and {Gibson}}{2004}]{Fan2004}
\begin{barticle}
\bauthor{\bsnm{{Fan}}, \binits{Y.}}, \bauthor{\bsnm{{Gibson}}, \binits{S.E.}}:
\byear{2004},
\batitle{Numerical simulations of three-dimensional coronal magnetic fields
  resulting from the emergence of twisted magnetic flux tubes}.
\bjtitle{\apjl}
\bvolume{609},
\bfpage{1123}\,--\,\blpage{1133}.
doi:\doiurl{10.1086/421238}.
\end{barticle}
\endbibitem

\bibitem[\protect\citeauthoryear{Fletcher and Hudson}{2001}]{Fletcher2001b}
\begin{barticle}
\bauthor{\bsnm{Fletcher}, \binits{L.}}, \bauthor{\bsnm{Hudson}, \binits{H.}}:
\byear{2001},
\batitle{The magnetic structure and generation of euv flare ribbons}.
\bjtitle{\solphys}
\bvolume{204},
\bfpage{69}\,--\,\blpage{89}.
\end{barticle}
\endbibitem

\bibitem[\protect\citeauthoryear{{Forbes} and {Priest}}{1995}]{Forbes1995}
\begin{barticle}
\bauthor{\bsnm{{Forbes}}, \binits{T.G.}}, \bauthor{\bsnm{{Priest}},
  \binits{E.R.}}:
\byear{1995},
\batitle{{Photospheric Magnetic Field Evolution and Eruptive Flares}}.
\bjtitle{\apj}
\bvolume{446},
\bfpage{377}.
doi:\doiurl{10.1086/175797}.
\end{barticle}
\endbibitem

\bibitem[\protect\citeauthoryear{{Gibson} and {Fan}}{2008}]{Gibson2008}
\begin{barticle}
\bauthor{\bsnm{{Gibson}}, \binits{S.E.}}, \bauthor{\bsnm{{Fan}}, \binits{Y.}}:
\byear{2008},
\batitle{{Partially ejected flux ropes: Implications for interplanetary coronal
  mass ejections}}.
\bjtitle{\jgrsp}
\bvolume{113},
\bfpage{9103}.
doi:\doiurl{10.1029/2008JA013151}.
\end{barticle}
\endbibitem

\bibitem[\protect\citeauthoryear{{Gopalswamy}
  \textit{et~al.}}{2001}]{Gopalswamy2001}
\begin{barticle}
\bauthor{\bsnm{{Gopalswamy}}, \binits{N.}}, \bauthor{\bsnm{{Yashiro}},
  \binits{S.}}, \bauthor{\bsnm{{Kaiser}}, \binits{M.L.}},
  \bauthor{\bsnm{{Howard}}, \binits{R.A.}}, \bauthor{\bsnm{{Bougeret}},
  \binits{J.}}:
\byear{2001},
\batitle{{Radio Signatures of Coronal Mass Ejection Interaction: Coronal Mass
  Ejection Cannibalism?}}
\bjtitle{\apjl}
\bvolume{548},
\bfpage{L91}\,--\,\blpage{L94}.
doi:\doiurl{10.1086/318939}.
\end{barticle}
\endbibitem

\bibitem[\protect\citeauthoryear{{Gopalswamy}
  \textit{et~al.}}{2010}]{Gopalswamy2010}
\begin{botherref}
\oauthor{\bsnm{{Gopalswamy}}, \binits{N.}}, \oauthor{\bsnm{{Akiyama}},
  \binits{S.}}, \oauthor{\bsnm{{Yashiro}}, \binits{S.}},
  \oauthor{\bsnm{{M{\"a}kel{\"a}}}, \binits{P.}}:
2010,
{Coronal Mass Ejections from Sunspot and Non-Sunspot Regions}.
In: {S.~S.~Hasan \& R.~J.~Rutten} (ed.)
\textit{Magnetic Coupling between the Interior and Atmosphere of the Sun},
289\,--\,307.
doi:\doiurl{10.1007/978}.
\end{botherref}
\endbibitem

\bibitem[\protect\citeauthoryear{{Gosling}}{1990}]{Gosling1990}
\begin{bchapter}
\bauthor{\bsnm{{Gosling}}, \binits{J.T.}}:
\byear{1990},
\bctitle{Coronal mass ejections and magnetic flux ropes in interplanetary
  space}.
In: \beditor{\bsnm{Russel}, \binits{C.T.}}, \beditor{\bsnm{Priest},
  \binits{E.R.}}, \beditor{\bsnm{Lee}, \binits{L.C.}} (eds.)
\bbtitle{Physics of Magnetic Flux Ropes},
\bsertitle{Geophys. Monographs}
\bseriesno{58},
\bpublisher{AGU}, \blocation{???},
\bfpage{343}\,--\,\blpage{364}.
\end{bchapter}
\endbibitem

\bibitem[\protect\citeauthoryear{{Green} \textit{et~al.}}{2002}]{Green2002}
\begin{barticle}
\bauthor{\bsnm{{Green}}, \binits{L.M.}}, \bauthor{\bsnm{{L{\'o}pez fuentes}},
  \binits{M.C.}}, \bauthor{\bsnm{{Mandrini}}, \binits{C.H.}},
  \bauthor{\bsnm{{D{\'e}moulin}}, \binits{P.}}, \bauthor{\bsnm{{Van
  Driel-Gesztelyi}}, \binits{L.}}, \bauthor{\bsnm{{Culhane}}, \binits{J.L.}}:
\byear{2002},
\batitle{The magnetic helicity budget of a cme-prolific active region}.
\bjtitle{\solphys}
\bvolume{208},
\bfpage{43}\,--\,\blpage{68}.
\end{barticle}
\endbibitem

\bibitem[\protect\citeauthoryear{{Gulisano}
  \textit{et~al.}}{2005}]{Gulisano2005}
\begin{barticle}
\bauthor{\bsnm{{Gulisano}}, \binits{A.M.}}, \bauthor{\bsnm{{Dasso}},
  \binits{S.}}, \bauthor{\bsnm{{Mandrini}}, \binits{C.H.}},
  \bauthor{\bsnm{{D{\'e}moulin}}, \binits{P.}}:
\byear{2005},
\batitle{{Magnetic clouds: A statistical study of magnetic helicity}}.
\bjtitle{Journal of Atmospheric and Solar-Terrestrial Physics}
\bvolume{67},
\bfpage{1761}\,--\,\blpage{1766}.
doi:\doiurl{10.1016/j.jastp.2005.02.026}.
\end{barticle}
\endbibitem

\bibitem[\protect\citeauthoryear{{Hirayama}}{1974}]{Hirayama1974}
\begin{barticle}
\bauthor{\bsnm{{Hirayama}}, \binits{T.}}:
\byear{1974},
\batitle{Theoretical model of flares and prominences. i: Evaporating flare
  model}.
\bjtitle{\solphys}
\bvolume{34},
\bfpage{323}\,--\,\blpage{338}.
\end{barticle}
\endbibitem

\bibitem[\protect\citeauthoryear{Hu and Sonnerup}{2001}]{Hu2001}
\begin{barticle}
\bauthor{\bsnm{Hu}, \binits{Q.}}, \bauthor{\bsnm{Sonnerup},
  \binits{B.U.{\"O}.}}:
\byear{2001},
\batitle{Reconstruction of magnetic flux ropes in the solar wind}.
\bjtitle{\grl}
\bvolume{28},
\bfpage{467}.
\end{barticle}
\endbibitem

\bibitem[\protect\citeauthoryear{{Jing} \textit{et~al.}}{2007}]{Jing2007}
\begin{barticle}
\bauthor{\bsnm{{Jing}}, \binits{J.}}, \bauthor{\bsnm{{Lee}}, \binits{J.}},
  \bauthor{\bsnm{{Liu}}, \binits{C.}}, \bauthor{\bsnm{{Gary}}, \binits{D.E.}},
  \bauthor{\bsnm{{Wang}}, \binits{H.}}:
\byear{2007},
\batitle{{Hard X-Ray Intensity Distribution along H{$\alpha$} Ribbons}}.
\bjtitle{\apjl}
\bvolume{664},
\bfpage{L127}\,--\,\blpage{L130}.
doi:\doiurl{10.1086/520812}.
\end{barticle}
\endbibitem

\bibitem[\protect\citeauthoryear{{Kazachenko}
  \textit{et~al.}}{2009}]{Kazachenko2009}
\begin{barticle}
\bauthor{\bsnm{{Kazachenko}}, \binits{M.D.}}, \bauthor{\bsnm{{Canfield}},
  \binits{R.C.}}, \bauthor{\bsnm{{Longcope}}, \binits{D.W.}},
  \bauthor{\bsnm{{Qiu}}, \binits{J.}}, \bauthor{\bsnm{{Des Jardins}},
  \binits{A.}}, \bauthor{\bsnm{{Nightingale}}, \binits{R.W.}}:
\byear{2009},
\batitle{{Sunspot Rotation, Flare Energetics, and Flux Rope Helicity: The
  Eruptive Flare on 2005 May 13}}.
\bjtitle{\apj}
\bvolume{704},
\bfpage{1146}\,--\,\blpage{1158}.
doi:\doiurl{10.1088/0004-637X/704/2/1146}.
\end{barticle}
\endbibitem

\bibitem[\protect\citeauthoryear{{Kazachenko}
  \textit{et~al.}}{2010}]{Kazachenko2010}
\begin{barticle}
\bauthor{\bsnm{{Kazachenko}}, \binits{M.D.}}, \bauthor{\bsnm{{Canfield}},
  \binits{R.C.}}, \bauthor{\bsnm{{Longcope}}, \binits{D.W.}},
  \bauthor{\bsnm{{Qiu}}, \binits{J.}}:
\byear{2010},
\batitle{{Sunspot Rotation, Flare Energetics, and Flux Rope Helicity: The
  Halloween Flare on 2003 October 28}}.
\bjtitle{\apj}
\bvolume{722},
\bfpage{1539}\,--\,\blpage{1546}.
doi:\doiurl{10.1088/0004-637X/722/2/1539}.
\end{barticle}
\endbibitem

\bibitem[\protect\citeauthoryear{{Kopp} and {Pneuman}}{1976}]{Kopp1976}
\begin{barticle}
\bauthor{\bsnm{{Kopp}}, \binits{R.A.}}, \bauthor{\bsnm{{Pneuman}},
  \binits{G.W.}}:
\byear{1976},
\batitle{Magnetic reconnection in the corona and the loop prominence
  phenomenon}.
\bjtitle{\solphys}
\bvolume{50},
\bfpage{85}\,--\,\blpage{98}.
\end{barticle}
\endbibitem

\bibitem[\protect\citeauthoryear{{Larson} \textit{et~al.}}{1997}]{Larson1997}
\begin{barticle}
\bauthor{\bsnm{{Larson}}, \binits{D.E.}}, \bauthor{\bsnm{{Lin}},
  \binits{R.P.}}, \bauthor{\bsnm{{McTiernan}}, \binits{J.M.}},
  \bauthor{\bsnm{{McFadden}}, \binits{J.P.}}, \bauthor{\bsnm{{Ergun}},
  \binits{R.E.}}, \bauthor{\bsnm{{McCarthy}}, \binits{M.}},
  \bauthor{\bsnm{{R{\` e}me}}, \binits{H.}}, \bauthor{\bsnm{{Sanderson}},
  \binits{T.R.}}, \bauthor{\bsnm{{Kaiser}}, \binits{M.}},
  \bauthor{\bsnm{{Lepping}}, \binits{R.P.}}, \bauthor{\bsnm{{Mazur}},
  \binits{J.}}:
\byear{1997},
\batitle{Tracing the topology of the october 18-20, 1995, magnetic cloud with
  $0.1-10^{2}$ kev electrons}.
\bjtitle{\grl}
\bvolume{24},
\bfpage{1911}\,--\,\blpage{1914}.
\end{barticle}
\endbibitem

\bibitem[\protect\citeauthoryear{{Leamon}, {Canfield}, and
  {Pevtsov}}{2002}]{Leamon.Canfield.Pevtsov2002_rcc}
\begin{barticle}
\bauthor{\bsnm{{Leamon}}, \binits{R.J.}}, \bauthor{\bsnm{{Canfield}},
  \binits{R.C.}}, \bauthor{\bsnm{{Pevtsov}}, \binits{A.A.}}:
\byear{2002},
\batitle{{Properties of Magnetic Clouds and Geomagnetic Storms associated
  Eruption of Coronal Sigmoids}}.
\bjtitle{\jgr}
\bvolume{107},
\bfpage{1234}.
\end{barticle}
\endbibitem

\bibitem[\protect\citeauthoryear{{Leamon} \textit{et~al.}}{2004}]{Leamon2004}
\begin{barticle}
\bauthor{\bsnm{{Leamon}}, \binits{R.J.}}, \bauthor{\bsnm{{Canfield}},
  \binits{R.C.}}, \bauthor{\bsnm{{Jones}}, \binits{S.L.}},
  \bauthor{\bsnm{{Lambkin}}, \binits{K.}}, \bauthor{\bsnm{{Lundberg}},
  \binits{B.J.}}, \bauthor{\bsnm{{Pevtsov}}, \binits{A.A.}}:
\byear{2004},
\batitle{Helicity of magnetic clouds and their associated active regions}.
\bjtitle{\jgr}
\bvolume{109},
\bfpage{5106}.
doi:\doiurl{10.1029/2003JA010324}.
\end{barticle}
\endbibitem

\bibitem[\protect\citeauthoryear{Leka \textit{et~al.}}{1996}]{Leka1996}
\begin{barticle}
\bauthor{\bsnm{Leka}, \binits{K.D.}}, \bauthor{\bsnm{Canfield}, \binits{R.C.}},
  \bauthor{\bsnm{McClymont}, \binits{A.N.}}, \bauthor{\bparticle{Van~{D}riel
  }\bsnm{{G}esztelyi}, \binits{L.}}:
\byear{1996},
\batitle{Evidence for current-carrying emerging flux}.
\bjtitle{\apjl}
\bvolume{462},
\bfpage{547}\,--\,\blpage{560}.
\end{barticle}
\endbibitem

\bibitem[\protect\citeauthoryear{{Lepping} \textit{et~al.}}{2001}]{Lepping2001}
\begin{barticle}
\bauthor{\bsnm{{Lepping}}, \binits{R.P.}}, \bauthor{\bsnm{{Berdichevsky}},
  \binits{D.B.}}, \bauthor{\bsnm{{Burlaga}}, \binits{L.F.}},
  \bauthor{\bsnm{{Lazarus}}, \binits{A.J.}}, \bauthor{\bsnm{{Kasper}},
  \binits{J.}}, \bauthor{\bsnm{{Desch}}, \binits{M.D.}}, \bauthor{\bsnm{{Wu}},
  \binits{C.}}, \bauthor{\bsnm{{Reames}}, \binits{D.V.}},
  \bauthor{\bsnm{{Singer}}, \binits{H.J.}}, \bauthor{\bsnm{{Smith}},
  \binits{C.W.}}, \bauthor{\bsnm{{Ackerson}}, \binits{K.L.}}:
\byear{2001},
\batitle{{The Bastille day Magnetic Clouds and Upstream Shocks: Near-Earth
  Interplanetary Observations}}.
\bjtitle{\solphys}
\bvolume{204},
\bfpage{285}\,--\,\blpage{303}.
doi:\doiurl{10.1023/A:1014264327855}.
\end{barticle}
\endbibitem

\bibitem[\protect\citeauthoryear{{Li} \textit{et~al.}}{2010}]{Li2010}
\begin{botherref}
\oauthor{\bsnm{{Li}}, \binits{Y.}}, \oauthor{\bsnm{{Luhman}}, \binits{J.G.}},
  \oauthor{\bsnm{{Lynch}}, \binits{B.J.}}, \oauthor{\bsnm{{Kilpua}},
  \binits{E.}}:
2010,
{Cyclic Reversal of Magnetic Cloud Poloidal Field}.
\textit{\solphys}.
\end{botherref}
\endbibitem

\bibitem[\protect\citeauthoryear{{Lin}, {Raymond}, and {van
  Ballegooijen}}{2004}]{Lin2004}
\begin{barticle}
\bauthor{\bsnm{{Lin}}, \binits{J.}}, \bauthor{\bsnm{{Raymond}}, \binits{J.C.}},
  \bauthor{\bsnm{{van Ballegooijen}}, \binits{A.A.}}:
\byear{2004},
\batitle{{The Role of Magnetic Reconnection in the Observable Features of Solar
  Eruptions}}.
\bjtitle{\apj}
\bvolume{602},
\bfpage{422}\,--\,\blpage{435}.
doi:\doiurl{10.1086/380900}.
\end{barticle}
\endbibitem

\bibitem[\protect\citeauthoryear{{Liu} \textit{et~al.}}{2007}]{Liu2007}
\begin{barticle}
\bauthor{\bsnm{{Liu}}, \binits{C.}}, \bauthor{\bsnm{{Lee}}, \binits{J.}},
  \bauthor{\bsnm{{Yurchyshyn}}, \binits{V.}}, \bauthor{\bsnm{{Deng}},
  \binits{N.}}, \bauthor{\bsnm{{Cho}}, \binits{K.s.}},
  \bauthor{\bsnm{{Karlick{\'y}}}, \binits{M.}}, \bauthor{\bsnm{{Wang}},
  \binits{H.}}:
\byear{2007},
\batitle{{The Eruption from a Sigmoidal Solar Active Region on 2005 May 13}}.
\bjtitle{\apj}
\bvolume{669},
\bfpage{1372}\,--\,\blpage{1381}.
doi:\doiurl{10.1086/521644}.
\end{barticle}
\endbibitem

\bibitem[\protect\citeauthoryear{{Liu}, {Zhang}, and {Zhang}}{2008}]{Liu2008}
\begin{barticle}
\bauthor{\bsnm{{Liu}}, \binits{J.}}, \bauthor{\bsnm{{Zhang}}, \binits{Y.}},
  \bauthor{\bsnm{{Zhang}}, \binits{H.}}:
\byear{2008},
\batitle{{Relationship between Powerful Flares and Dynamic Evolution of the
  Magnetic Field at the Solar Surface}}.
\bjtitle{\solphys}
\bvolume{248},
\bfpage{67}\,--\,\blpage{84}.
doi:\doiurl{10.1007/s11207-008-9149-0}.
\end{barticle}
\endbibitem

\bibitem[\protect\citeauthoryear{{Longcope}}{1996}]{Longcope1996d}
\begin{barticle}
\bauthor{\bsnm{{Longcope}}, \binits{D.W.}}:
\byear{1996},
\batitle{{Topology and Current Ribbons: A Model for Current, Reconnection and
  Flaring in a Complex, Evolving Corona}}.
\bjtitle{\solphys}
\bvolume{169},
\bfpage{91}\,--\,\blpage{121}.
doi:\doiurl{10.1007/BF00153836}.
\end{barticle}
\endbibitem

\bibitem[\protect\citeauthoryear{{Longcope}}{2001}]{Longcope2001b}
\begin{barticle}
\bauthor{\bsnm{{Longcope}}, \binits{D.W.}}:
\byear{2001},
\batitle{{Separator current sheets: Generic features in minimum-energy magnetic
  fields subject to flux constraints}}.
\bjtitle{Physics of Plasmas}
\bvolume{8},
\bfpage{5277}\,--\,\blpage{5290}.
doi:\doiurl{10.1063/1.1418431}.
\end{barticle}
\endbibitem

\bibitem[\protect\citeauthoryear{{Longcope} and {Magara}}{2004}]{Longcope2004}
\begin{barticle}
\bauthor{\bsnm{{Longcope}}, \binits{D.W.}}, \bauthor{\bsnm{{Magara}},
  \binits{T.}}:
\byear{2004},
\batitle{{A Comparison of the Minimum Current Corona to a Magnetohydrodynamic
  Simulation of Quasi-Static Coronal Evolution}}.
\bjtitle{\apj}
\bvolume{608},
\bfpage{1106}\,--\,\blpage{1123}.
doi:\doiurl{10.1086/420780}.
\end{barticle}
\endbibitem

\bibitem[\protect\citeauthoryear{{Longcope}
  \textit{et~al.}}{2010}]{Longcope2010}
\begin{botherref}
\oauthor{\bsnm{{Longcope}}, \binits{D.W.}}, \oauthor{\bsnm{{DesJardins}},
  \binits{A.C.}}, \oauthor{\bsnm{{Carranza-Fulmer}}, \binits{T.}},
  \oauthor{\bsnm{{Qiu}}, \binits{J.}}:
2010,
{A Quantitative Model of Energy Release and Heating by Time-dependent,
  Localized Reconnection in a Flare with a Thermal Loop-top X-ray Source}.
\textit{\solphys}.
\end{botherref}
\endbibitem

\bibitem[\protect\citeauthoryear{{Longcope}
  \textit{et~al.}}{2007}]{Longcope2007}
\begin{barticle}
\bauthor{\bsnm{{Longcope}}, \binits{D.W.}}, \bauthor{\bsnm{{Beveridge}},
  \binits{C.}}, \bauthor{\bsnm{{Qiu}}, \binits{J.}},
  \bauthor{\bsnm{{Ravindra}}, \binits{B.}}, \bauthor{\bsnm{{Barnes}},
  \binits{G.}}, \bauthor{\bsnm{{Dasso}}, \binits{S.}}:
\byear{2007},
\batitle{{Modeling and Measuring the Flux Reconnected and Ejected by the
  Two-Ribbon Flare/CME Event on 7 November 2004}}.
\bjtitle{\solphys}
\bvolume{244},
\bfpage{45}\,--\,\blpage{73}.
doi:\doiurl{10.1007/s11207-007-0330-7}.
\end{barticle}
\endbibitem


\bibitem[\protect\citeauthoryear{{Longcope}, {Barnes} and {Beveridge}}{2009}]{Longcope2009}
\begin{barticle}
\bauthor{\bsnm{{Longcope}}, \binits{D.W.}},  \bauthor{\bsnm{{Barnes}}, \binits{G.}}, 
\bauthor{\bsnm{{Beveridge}},  \binits{C.}}:
\byear{2009},
\batitle{{Effects of Partitioning and Extrapolation on the Connectivity of Potential Magnetic Fields}}.
\bjtitle{\apj}
\bvolume{693},
\bfpage{97}\,--\,\blpage{111}.
doi:\doiurl{10.1088/0004-637X/693/1/97}.
\end{barticle}
\endbibitem

\bibitem[\protect\citeauthoryear{{Low}}{1994}]{Low1994}
\begin{barticle}
\bauthor{\bsnm{{Low}}, \binits{B.C.}}:
\byear{1994},
\batitle{Magnetohydrodynamic processes in the solar corona: Flares, coronal
  mass ejections, and magnetic helicity}.
\bjtitle{Physics of Plasmas}
\bvolume{1},
\bfpage{1684}\,--\,\blpage{1690}.
\end{barticle}
\endbibitem

\bibitem[\protect\citeauthoryear{{Luoni} \textit{et~al.}}{2005}]{Luoni2005}
\begin{barticle}
\bauthor{\bsnm{{Luoni}}, \binits{M.L.}}, \bauthor{\bsnm{{Mandrini}},
  \binits{C.H.}}, \bauthor{\bsnm{{Dasso}}, \binits{S.}}, \bauthor{\bsnm{{van
  Driel-Gesztelyi}}, \binits{L.}}, \bauthor{\bsnm{{D{\'e}moulin}},
  \binits{P.}}:
\byear{2005},
\batitle{{Tracing magnetic helicity from the solar corona to the interplanetary
  space}}.
\bjtitle{Journal of Atmospheric and Solar-Terrestrial Physics}
\bvolume{67},
\bfpage{1734}\,--\,\blpage{1743}.
doi:\doiurl{10.1016/j.jastp.2005.07.003}.
\end{barticle}
\endbibitem

\bibitem[\protect\citeauthoryear{{Lynch} \textit{et~al.}}{2005}]{Lynch2005}
\begin{barticle}
\bauthor{\bsnm{{Lynch}}, \binits{B.J.}}, \bauthor{\bsnm{{Gruesbeck}},
  \binits{J.R.}}, \bauthor{\bsnm{{Zurbuchen}}, \binits{T.H.}},
  \bauthor{\bsnm{{Antiochos}}, \binits{S.K.}}:
\byear{2005},
\batitle{{Solar cycle-dependent helicity transport by magnetic clouds}}.
\bjtitle{\jgrsp}
\bvolume{110},
\bfpage{8107}.
doi:\doiurl{10.1029/2005JA011137}.
\end{barticle}
\endbibitem

\bibitem[\protect\citeauthoryear{{Mackay} and {van
  Ballegooijen}}{2006}]{Mackay2006}
\begin{barticle}
\bauthor{\bsnm{{Mackay}}, \binits{D.H.}}, \bauthor{\bsnm{{van Ballegooijen}},
  \binits{A.A.}}:
\byear{2006},
\batitle{{Models of the Large-Scale Corona. I. Formation, Evolution, and
  Liftoff of Magnetic Flux Ropes}}.
\bjtitle{\apj}
\bvolume{641},
\bfpage{577}\,--\,\blpage{589}.
doi:\doiurl{10.1086/500425}.
\end{barticle}
\endbibitem

\bibitem[\protect\citeauthoryear{{Mandrini}
  \textit{et~al.}}{2005}]{Mandrini2005}
\begin{barticle}
\bauthor{\bsnm{{Mandrini}}, \binits{C.H.}}, \bauthor{\bsnm{{Pohjolainen}},
  \binits{S.}}, \bauthor{\bsnm{{Dasso}}, \binits{S.}}, \bauthor{\bsnm{{Green}},
  \binits{L.M.}}, \bauthor{\bsnm{{D{\'e}moulin}}, \binits{P.}},
  \bauthor{\bsnm{{van Driel-Gesztelyi}}, \binits{L.}},
  \bauthor{\bsnm{{Copperwheat}}, \binits{C.}}, \bauthor{\bsnm{{Foley}},
  \binits{C.}}:
\byear{2005},
\batitle{{Interplanetary flux rope ejected from an X-ray bright point. The
  smallest magnetic cloud source-region ever observed}}.
\bjtitle{\aap}
\bvolume{434},
\bfpage{725}\,--\,\blpage{740}.
doi:\doiurl{10.1051/0004-6361:20041079}.
\end{barticle}
\endbibitem

\bibitem[\protect\citeauthoryear{{Mandrini}
  \textit{et~al.}}{2006}]{Mandrini2006}
\begin{barticle}
\bauthor{\bsnm{{Mandrini}}, \binits{C.H.}}, \bauthor{\bsnm{{Demoulin}},
  \binits{P.}}, \bauthor{\bsnm{{Schmieder}}, \binits{B.}},
  \bauthor{\bsnm{{Deluca}}, \binits{E.E.}}, \bauthor{\bsnm{{Pariat}},
  \binits{E.}}, \bauthor{\bsnm{{Uddin}}, \binits{W.}}:
\byear{2006},
\batitle{{Companion Event and Precursor of the X17 Flare on 28 October 2003}}.
\bjtitle{\solphys}
\bvolume{238},
\bfpage{293}\,--\,\blpage{312}.
doi:\doiurl{10.1007/s11207-006-0205-3}.
\end{barticle}
\endbibitem

\bibitem[\protect\citeauthoryear{{Marubashi}}{1986}]{Marubashi1986}
\begin{barticle}
\bauthor{\bsnm{{Marubashi}}, \binits{K.}}:
\byear{1986},
\batitle{{Structure of the interplanetary magnetic clouds and their solar
  origins}}.
\bjtitle{Advances in Space Research}
\bvolume{6},
\bfpage{335}\,--\,\blpage{338}.
doi:\doiurl{10.1016/0273-1177(86)90172-9}.
\end{barticle}
\endbibitem

\bibitem[\protect\citeauthoryear{{Masuda}, {Kosugi}, and
  {Hudson}}{2001}]{Masuda2001}
\begin{barticle}
\bauthor{\bsnm{{Masuda}}, \binits{S.}}, \bauthor{\bsnm{{Kosugi}}, \binits{T.}},
  \bauthor{\bsnm{{Hudson}}, \binits{H.S.}}:
\byear{2001},
\batitle{{A Hard X-ray Two-Ribbon Flare Observed with Yohkoh/HXT}}.
\bjtitle{\solphys}
\bvolume{204},
\bfpage{55}\,--\,\blpage{67}.
doi:\doiurl{10.1023/A:1014230629731}.
\end{barticle}
\endbibitem

\bibitem[\protect\citeauthoryear{{Mewe}, {Gronenschild}, and {van den
  Oord}}{1985}]{Mewe1985}
\begin{barticle}
\bauthor{\bsnm{{Mewe}}, \binits{R.}}, \bauthor{\bsnm{{Gronenschild}},
  \binits{E.H.B.M.}}, \bauthor{\bsnm{{van den Oord}}, \binits{G.H.J.}}:
\byear{1985},
\batitle{{Calculated X-radiation from optically thin plasmas. V}}.
\bjtitle{\aaps}
\bvolume{62},
\bfpage{197}\,--\,\blpage{254}.
\end{barticle}
\endbibitem

\bibitem[\protect\citeauthoryear{{Mulligan}, {Russell}, and
  {Luhmann}}{1998}]{Mulligan1998}
\begin{barticle}
\bauthor{\bsnm{{Mulligan}}, \binits{T.}}, \bauthor{\bsnm{{Russell}},
  \binits{C.T.}}, \bauthor{\bsnm{{Luhmann}}, \binits{J.G.}}:
\byear{1998},
\batitle{{Solar cycle evolution of the structure of magnetic clouds in the
  inner heliosphere}}.
\bjtitle{\grl}
\bvolume{25},
\bfpage{2959}\,--\,\blpage{2962}.
doi:\doiurl{10.1029/98GL01302}.
\end{barticle}
\endbibitem

\bibitem[\protect\citeauthoryear{{Nindos}, {Zhang}, and
  {Zhang}}{2003}]{Nindos2003}
\begin{barticle}
\bauthor{\bsnm{{Nindos}}, \binits{A.}}, \bauthor{\bsnm{{Zhang}}, \binits{J.}},
  \bauthor{\bsnm{{Zhang}}, \binits{H.}}:
\byear{2003},
\batitle{The magnetic helicity budget of solar active regions and coronal mass
  ejections}.
\bjtitle{\apjl}
\bvolume{594},
\bfpage{1033}\,--\,\blpage{1048}.
doi:\doiurl{10.1086/377126}.
\end{barticle}
\endbibitem

\bibitem[\protect\citeauthoryear{{November} and {Simon}}{1988}]{November1988}
\begin{barticle}
\bauthor{\bsnm{{November}}, \binits{L.J.}}, \bauthor{\bsnm{{Simon}},
  \binits{G.W.}}:
\byear{1988},
\batitle{{Precise proper-motion measurement of solar granulation}}.
\bjtitle{\apj}
\bvolume{333},
\bfpage{427}\,--\,\blpage{442}.
doi:\doiurl{10.1086/166758}.
\end{barticle}
\endbibitem

\bibitem[\protect\citeauthoryear{Pariat, D{\'e}moulin, and
  Berger}{2005}]{Pariat2005}
\begin{barticle}
\bauthor{\bsnm{Pariat}, \binits{E.}}, \bauthor{\bsnm{D{\'e}moulin},
  \binits{P.}}, \bauthor{\bsnm{Berger}, \binits{M.A.}}:
\byear{2005},
\batitle{Photospheric flux density of magnetic helicity}.
\bjtitle{\aap}
\bvolume{439},
\bfpage{1191}\,--\,\blpage{1203}.
\end{barticle}
\endbibitem

\bibitem[\protect\citeauthoryear{{Pevtsov} and
  {Balasubramaniam}}{2003}]{Pevtsov2003d}
\begin{barticle}
\bauthor{\bsnm{{Pevtsov}}, \binits{A.A.}}, \bauthor{\bsnm{{Balasubramaniam}},
  \binits{K.S.}}:
\byear{2003},
\batitle{{Helicity patterns on the sun}}.
\bjtitle{Advances in Space Research}
\bvolume{32},
\bfpage{1867}\,--\,\blpage{1874}.
doi:\doiurl{10.1016/S0273-1177(03)90620-X}.
\end{barticle}
\endbibitem

\bibitem[\protect\citeauthoryear{Poletto and Kopp}{1986}]{Poletto1986}
\begin{bchapter}
\bauthor{\bsnm{Poletto}, \binits{G.}}, \bauthor{\bsnm{Kopp}, \binits{R.A.}}:
\byear{1986},
\bctitle{Macroscopic electric fields during two-ribbon flares}.
In: \beditor{\bsnm{Neidig}, \binits{D.F.}} (ed.)
\bbtitle{The Lower Atmospheres of Solar Flares},
\bpublisher{National Solar Observatory}, \blocation{???},
\bfpage{453}\,--\,\blpage{465}.
\end{bchapter}
\endbibitem

\bibitem[\protect\citeauthoryear{{Qiu} and {Gary}}{2003}]{Qiu2003}
\begin{barticle}
\bauthor{\bsnm{{Qiu}}, \binits{J.}}, \bauthor{\bsnm{{Gary}}, \binits{D.E.}}:
\byear{2003},
\batitle{{Flare-related Magnetic Anomaly with a Sign Reversal}}.
\bjtitle{\apj}
\bvolume{599},
\bfpage{615}\,--\,\blpage{625}.
doi:\doiurl{10.1086/379146}.
\end{barticle}
\endbibitem

\bibitem[\protect\citeauthoryear{{Qiu} \textit{et~al.}}{2007}]{Qiu2007}
\begin{barticle}
\bauthor{\bsnm{{Qiu}}, \binits{J.}}, \bauthor{\bsnm{{Hu}}, \binits{Q.}},
  \bauthor{\bsnm{{Howard}}, \binits{T.A.}}, \bauthor{\bsnm{{Yurchyshyn}},
  \binits{V.B.}}:
\byear{2007},
\batitle{{On the Magnetic Flux Budget in Low-Corona Magnetic Reconnection and
  Interplanetary Coronal Mass Ejections}}.
\bjtitle{\apj}
\bvolume{659},
\bfpage{758}\,--\,\blpage{772}.
doi:\doiurl{10.1086/512060}.
\end{barticle}
\endbibitem

\bibitem[\protect\citeauthoryear{{Ravindra} and
  {Howard}}{2010}]{Ravindra2010}
\begin{barticle}
\bauthor{\bsnm{{Ravindra}}, \binits{B.}}, \bauthor{\bsnm{{Howard}},
  \binits{T.~A.}}:
\byear{2010},
\batitle{{Comparison of energies between eruptive phenomena and magnetic field in AR 10930}}.
\bjtitle{\basi}
\bvolume{38},
\bfpage{L147}\,--\,\blpage{L163}.
doi:\doiurl{http://cdsads.u-strasbg.fr/abs/2010BASI...38..147R}.
\end{barticle}
\endbibitem

\bibitem[\protect\citeauthoryear{{R{\'e}gnier} and
  {Priest}}{2007}]{Regnier2007}
\begin{barticle}
\bauthor{\bsnm{{R{\'e}gnier}}, \binits{S.}}, \bauthor{\bsnm{{Priest}},
  \binits{E.R.}}:
\byear{2007},
\batitle{{Free Magnetic Energy in Solar Active Regions above the Minimum-Energy
  Relaxed State}}.
\bjtitle{\apjl}
\bvolume{669},
\bfpage{L53}\,--\,\blpage{L56}.
doi:\doiurl{10.1086/523269}.
\end{barticle}
\endbibitem

\bibitem[\protect\citeauthoryear{Rosner, Tucker, and Vaiana}{1978}]{Rosner1978}
\begin{barticle}
\bauthor{\bsnm{Rosner}, \binits{R.}}, \bauthor{\bsnm{Tucker}, \binits{W.H.}},
  \bauthor{\bsnm{Vaiana}, \binits{G.S.}}:
\byear{1978},
\batitle{Dynamics of the quiescent corona}.
\bjtitle{\apjl}
\bvolume{220},
\bfpage{643}\,--\,\blpage{655}.
\end{barticle}
\endbibitem

\bibitem[\protect\citeauthoryear{{Scherrer}
  \textit{et~al.}}{1995}]{Scherrer1995}
\begin{barticle}
\bauthor{\bsnm{{Scherrer}}, \binits{P.H.}}, \bauthor{\bsnm{{Bogart}},
  \binits{R.S.}}, \bauthor{\bsnm{{Bush}}, \binits{R.I.}},
  \bauthor{\bsnm{{Hoeksema}}, \binits{J.T.}}, \bauthor{\bsnm{{Kosovichev}},
  \binits{A.G.}}, \bauthor{\bsnm{{Schou}}, \binits{J.}},
  \bauthor{\bsnm{{Rosenberg}}, \binits{W.}}, \bauthor{\bsnm{{Springer}},
  \binits{L.}}, \bauthor{\bsnm{{Tarbell}}, \binits{T.D.}},
  \bauthor{\bsnm{{Title}}, \binits{A.}}, \bauthor{\bsnm{{Wolfson}},
  \binits{C.J.}}, \bauthor{\bsnm{{Zayer}}, \binits{I.}}, \bauthor{\bsnm{{MDI
  Engineering Team}}}:
\byear{1995},
\batitle{{The Solar Oscillations Investigation - Michelson Doppler Imager}}.
\bjtitle{\solphys}
\bvolume{162},
\bfpage{129}\,--\,\blpage{188}.
doi:\doiurl{10.1007/BF00733429}.
\end{barticle}
\endbibitem

\bibitem[\protect\citeauthoryear{{Schrijver}
  \textit{et~al.}}{2006}]{Schrijver2006}
\begin{barticle}
\bauthor{\bsnm{{Schrijver}}, \binits{C.J.}}, \bauthor{\bsnm{{Derosa}},
  \binits{M.L.}}, \bauthor{\bsnm{{Metcalf}}, \binits{T.R.}},
  \bauthor{\bsnm{{Liu}}, \binits{Y.}}, \bauthor{\bsnm{{McTiernan}},
  \binits{J.}}, \bauthor{\bsnm{{R{\'e}gnier}}, \binits{S.}},
  \bauthor{\bsnm{{Valori}}, \binits{G.}}, \bauthor{\bsnm{{Wheatland}},
  \binits{M.S.}}, \bauthor{\bsnm{{Wiegelmann}}, \binits{T.}}:
\byear{2006},
\batitle{Nonlinear force-free modeling of coronal magnetic fields part i: A
  quantitative comparison of methods}.
\bjtitle{\solphys}
\bvolume{235},
\bfpage{161}\,--\,\blpage{190}.
doi:\doiurl{10.1007/s11207-006-0068-7}.
\end{barticle}
\endbibitem

\bibitem[\protect\citeauthoryear{{Sol.Physics special
  edition}}{2001}]{Bast2001}
\begin{barticle}
\bauthor{\bsnm{{Sol.Physics special edition}}}:
\byear{2001},
\batitle{{2000 Bastille Day Flare Event}}.
\bjtitle{\solphys}
\bvolume{204},
\bfpage{1}\,--\,\blpage{436}.
doi:\doiurl{10.1086/420780}.
\end{barticle}
\endbibitem

\bibitem[\protect\citeauthoryear{{Sturrock}}{1968}]{Sturrock1968}
\begin{botherref}
\oauthor{\bsnm{{Sturrock}}, \binits{P.A.}}:
1968,
A model of solar flares.
In: \textit{IAU Symp.\ 35: Structure and Development of Solar Active Regions},
471\,--\,479.
\end{botherref}
\endbibitem

\bibitem[\protect\citeauthoryear{{Su}, {Golub}, and {Van
  Ballegooijen}}{2007}]{Su2007}
\begin{barticle}
\bauthor{\bsnm{{Su}}, \binits{Y.}}, \bauthor{\bsnm{{Golub}}, \binits{L.}},
  \bauthor{\bsnm{{Van Ballegooijen}}, \binits{A.A.}}:
\byear{2007},
\batitle{{A Statistical Study of Shear Motion of the Footpoints in Two-Ribbon
  Flares}}.
\bjtitle{\apj}
\bvolume{655},
\bfpage{606}\,--\,\blpage{614}.
doi:\doiurl{10.1086/510065}.
\end{barticle}
\endbibitem

\bibitem[\protect\citeauthoryear{Vesecky, Antiochos, and
  Underwood}{1979}]{Vesecky1979}
\begin{barticle}
\bauthor{\bsnm{Vesecky}, \binits{J.F.}}, \bauthor{\bsnm{Antiochos},
  \binits{S.K.}}, \bauthor{\bsnm{Underwood}, \binits{J.H.}}:
\byear{1979},
\batitle{Numerical modeling of quasi-static loops. {I}. uniform energy input}.
\bjtitle{\apjl}
\bvolume{233},
\bfpage{987}.
\end{barticle}
\endbibitem

\bibitem[\protect\citeauthoryear{{Webb}
  \textit{et~al.}}{2000}]{Webb.etal2000_rcc}
\begin{barticle}
\bauthor{\bsnm{{Webb}}, \binits{D.F.}}, \bauthor{\bsnm{{Lepping}},
  \binits{R.P.}}, \bauthor{\bsnm{{Burlaga}}, \binits{L.F.}},
  \bauthor{\bsnm{{Deforest}}, \binits{C.E.}}, \bauthor{\bsnm{{Larson}},
  \binits{D.E.}}, \bauthor{\bsnm{{Martin}}, \binits{S.F.}},
  \bauthor{\bsnm{{Plunkett}}, \binits{S.P.}}, \bauthor{\bsnm{{Rust}},
  \binits{D.M.}}:
\byear{2000},
\batitle{{The origin and development of the May 1997 magnetic cloud}}.
\bjtitle{\jgr}
\bvolume{105},
\bfpage{27251}\,--\,\blpage{27260}.
\end{barticle}
\endbibitem

\bibitem[\protect\citeauthoryear{Welsch \textit{et~al.}}{2007}]{Welsch2007}
\begin{barticle}
\bauthor{\bsnm{Welsch}, \binits{B.T.}}, \bauthor{\bsnm{Abbett}, \binits{W.P.}},
  \bauthor{\bsnm{DeRosa}, \binits{M.L.}}, \bauthor{\bsnm{Fisher},
  \binits{G.H.}}, \bauthor{\bsnm{Georgoulis}, \binits{M.K.}},
  \bauthor{\bsnm{Kusano}, \binits{K.}}, \bauthor{\bsnm{Longcope},
  \binits{D.W.}}, \bauthor{\bsnm{Ravindra}, \binits{B.}},
  \bauthor{\bsnm{Schuck}, \binits{P.W.}}:
\byear{2007},
\batitle{Tests and comparisons of velocity inversion techniques}.
\bjtitle{\apjl}
\bvolume{670},
\bfpage{1434}\,--\,\blpage{1452}.
\end{barticle}
\endbibitem

\bibitem[\protect\citeauthoryear{{Woods}, {Kopp}, and
  {Chamberlin}}{2006}]{Woods2006}
\begin{barticle}
\bauthor{\bsnm{{Woods}}, \binits{T.N.}}, \bauthor{\bsnm{{Kopp}}, \binits{G.}},
  \bauthor{\bsnm{{Chamberlin}}, \binits{P.C.}}:
\byear{2006},
\batitle{{Contributions of the solar ultraviolet irradiance to the total solar
  irradiance during large flares}}.
\bjtitle{\jgrsp}
\bvolume{111},
\bfpage{10}.
doi:\doiurl{10.1029/2005JA011507}.
\end{barticle}
\endbibitem

\bibitem[\protect\citeauthoryear{{Yurchyshyn}, {Abramenko}, and
  {Tripathi}}{2009}]{Yurchyshyn2009}
\begin{barticle}
\bauthor{\bsnm{{Yurchyshyn}}, \binits{V.}}, \bauthor{\bsnm{{Abramenko}},
  \binits{V.}}, \bauthor{\bsnm{{Tripathi}}, \binits{D.}}:
\byear{2009},
\batitle{{Rotation of White-light Coronal Mass Ejection Structures as Inferred
  from LASCO Coronagraph}}.
\bjtitle{\apj}
\bvolume{705},
\bfpage{426}\,--\,\blpage{435}.
doi:\doiurl{10.1088/0004-637X/705/1/426}.
\end{barticle}
\endbibitem

\bibitem[\protect\citeauthoryear{{Yurchyshyn}, {Hu}, and
  {Abramenko}}{2005}]{Yurchyshyn2005}
\begin{barticle}
\bauthor{\bsnm{{Yurchyshyn}}, \binits{V.}}, \bauthor{\bsnm{{Hu}}, \binits{Q.}},
  \bauthor{\bsnm{{Abramenko}}, \binits{V.}}:
\byear{2005},
\batitle{{Structure of magnetic fields in NOAA active regions 0486 and 0501 and
  in the associated interplanetary ejecta}}.
\bjtitle{Space Weather}
\bvolume{3},
\bfpage{8}.
doi:\doiurl{10.1029/2004SW000124}.
\end{barticle}
\endbibitem

\bibitem[\protect\citeauthoryear{{Yurchyshyn}
  \textit{et~al.}}{2001}]{Yurchyshyn2001}
\begin{barticle}
\bauthor{\bsnm{{Yurchyshyn}}, \binits{V.B.}}, \bauthor{\bsnm{{Wang}},
  \binits{H.}}, \bauthor{\bsnm{{Goode}}, \binits{P.R.}},
  \bauthor{\bsnm{{Deng}}, \binits{Y.}}:
\byear{2001},
\batitle{{Orientation of the Magnetic Fields in Interplanetary Flux Ropes and
  Solar Filaments}}.
\bjtitle{\apj}
\bvolume{563},
\bfpage{381}\,--\,\blpage{388}.
doi:\doiurl{10.1086/323778}.
\end{barticle}
\endbibitem

\bibitem[\protect\citeauthoryear{{Yurchyshyn}
  \textit{et~al.}}{2006}]{Yurchyshyn2006}
\begin{barticle}
\bauthor{\bsnm{{Yurchyshyn}}, \binits{V.}}, \bauthor{\bsnm{{Liu}},
  \binits{C.}}, \bauthor{\bsnm{{Abramenko}}, \binits{V.}},
  \bauthor{\bsnm{{Krall}}, \binits{J.}}:
\byear{2006},
\batitle{{The May 13, 2005 Eruption: Observations, Data Analysis and
  Interpretation}}.
\bjtitle{\solphys}
\bvolume{239},
\bfpage{317}\,--\,\blpage{335}.
doi:\doiurl{10.1007/s11207-006-0177-3}.
\end{barticle}
\endbibitem

\bibitem[\protect\citeauthoryear{{Yurchyshyn}
  \textit{et~al.}}{2007}]{Yurchyshyn2007}
\begin{barticle}
\bauthor{\bsnm{{Yurchyshyn}}, \binits{V.}}, \bauthor{\bsnm{{Hu}}, \binits{Q.}},
  \bauthor{\bsnm{{Lepping}}, \binits{R.P.}}, \bauthor{\bsnm{{Lynch}},
  \binits{B.J.}}, \bauthor{\bsnm{{Krall}}, \binits{J.}}:
\byear{2007},
\batitle{{Orientations of LASCO Halo CMEs and their connection to the flux rope
  structure of interplanetary CMEs}}.
\bjtitle{Advances in Space Research}
\bvolume{40},
\bfpage{1821}\,--\,\blpage{1826}.
doi:\doiurl{10.1016/j.asr.2007.01.059}.
\end{barticle}
\endbibitem

\bibitem[\protect\citeauthoryear{{Zhang}, {Liu}, and {Zhang}}{2008}]{Zhang2008}
\begin{barticle}
\bauthor{\bsnm{{Zhang}}, \binits{Y.}}, \bauthor{\bsnm{{Liu}}, \binits{J.}},
  \bauthor{\bsnm{{Zhang}}, \binits{H.}}:
\byear{2008},
\batitle{{Relationship between Rotating Sunspots and Flares}}.
\bjtitle{\solphys}
\bvolume{247},
\bfpage{39}\,--\,\blpage{52}.
doi:\doiurl{10.1007/s11207-007-9089-0}.
\end{barticle}
\endbibitem

\bibitem[\protect\citeauthoryear{{Zhao} and {Hoeksema}}{1998}]{Zhao1998}
\begin{barticle}
\bauthor{\bsnm{{Zhao}}, \binits{X.P.}}, \bauthor{\bsnm{{Hoeksema}},
  \binits{J.T.}}:
\byear{1998},
\batitle{{Central axial field direction in magnetic clouds and its relation to
  southward interplanetary magnetic field events and dependence on disappearing
  solar filaments}}.
\bjtitle{\jgr}
\bvolume{103},
\bfpage{2077}.
doi:\doiurl{10.1029/97JA03234}.
\end{barticle}
\endbibitem

\end{thebibliography}


\end{article}

\end{document}